\documentclass[nofootinbib,eqsecnum,superscriptaddress,amsmath,amssymb,aps,11pt]{revtex4-2}

\usepackage{amsfonts}
\usepackage{amsthm} 
\usepackage{bbm} 
\usepackage{multirow}

\usepackage{tikz}
\usetikzlibrary{quantikz2}

\usepackage{hyperref} 

\DeclareMathOperator{\Tr}{Tr} 
\newtheorem{theorem}{Theorem}

\newcommand{\secref}[1]{Sec.~\ref{#1}}

\newcommand{\figref}[1]{Fig.~\ref{#1}}
\newcommand{\appref}[1]{Appendix~\ref{#1}}
\newcommand{\ftref}[1]{Footnote~\ref{#1}}

\tikzset{
uctrl/.style={
        draw,
        circle,line width = 0.1pt,
        inner sep=0.4ex,
        append after command={
            \pgfextra {\draw[fill,black] (0,0) circle (0.5ex) ;
                \fill[white] (0,0) -- (45:0.5ex) arc (45:225:0.5ex) -- cycle;}}
},
}
\DeclareExpandableDocumentCommand{\uctrl}{O{}{m}}{|[uctrl,#1]| \vqw{#2} \qw}

\DeclareMathAlphabet{\mathpzc}{OT1}{pzc}{m}{it} 

\begin{document}
\count\footins = 1000 

\title{Quantum circuit synthesis for the preparation of arbitrary and highly sparse mixed quantum states}


\author{Bo-Hung Chen}
\email{kenny81778189@gmail.com}
\affiliation{Graduate Institute of Electronics Engineering, National Taiwan University, Taipei 10617, Taiwan}
\affiliation{Center for Quantum Science and Engineering, National Taiwan University, Taipei 10617, Taiwan}
\affiliation{Physics Division, National Center for Theoretical Sciences, Taipei 10617, Taiwan}

\author{Dah-Wei Chiou}
\email{dwchiou@gmail.com}
\affiliation{Graduate Institute of Electronics Engineering, National Taiwan University, Taipei 10617, Taiwan}
\affiliation{Center for Quantum Science and Engineering, National Taiwan University, Taipei 10617, Taiwan}
\affiliation{Physics Division, National Center for Theoretical Sciences, Taipei 10617, Taiwan}

\author{Jie-Hong Roland Jiang}
\email{jhjiang@ntu.edu.tw}
\affiliation{Graduate Institute of Electronics Engineering, National Taiwan University, Taipei 10617, Taiwan}
\affiliation{Center for Quantum Science and Engineering, National Taiwan University, Taipei 10617, Taiwan}
\affiliation{Physics Division, National Center for Theoretical Sciences, Taipei 10617, Taiwan}
\affiliation{Department of Electrical Engineering, National Taiwan University, Taipei 10617, Taiwan}

\begin{abstract}
This paper addresses the challenge of preparing mixed quantum states---both arbitrary states in general and highly sparse ones in particular---an area that has received far less attention than the preparation of pure states. We present two classes of circuit-synthesis methods: one based on constructing the density matrix as a mixture of pure states and the other based on purification. To improve both preprocessing efficiency and the complexity of the resulting circuits, we propose a novel strategy based on the Cholesky decomposition, which offers significant advantages, especially when the target density matrix is highly or extremely sparse. Furthermore, by exploiting incomplete Cholesky decomposition with threshold dropping, we introduce a practical approach for constructing high-fidelity approximations of the target density matrix. This approach enables substantial reductions in circuit complexity at the cost of negligible or only mild fidelity loss.
\end{abstract}

\maketitle

\section{Introduction}\label{sec:introduction}
Quantum state preparation is a fundamental task in quantum computation and quantum information, underpinning a wide range of quantum algorithms and protocols \cite{nielsen2010quantum}. The ability to prepare quantum systems in prescribed states---pure or mixed---enables the controlled exploitation of superposition, entanglement, and interference, and thus plays a central role in realizing the advantages of quantum information processing.

Efficient preparation of pure quantum states is a prerequisite for many quantum algorithms, including quantum search \cite{ambainis2004quantum,giri2017review}, quantum simulation \cite{RevModPhys.86.153}, and quantum machine learning \cite{wittek2014quantum,biamonte2017quantum}. In quantum communication and cryptography \cite{jaeger2007quantum,wilde2013quantum,RevModPhys.74.145}, high-fidelity state preparation is essential for the reliable manipulation and transmission of quantum information. Accurate pure-state preparation is also critical in applications such as the quantum algorithm for solving linear systems \cite{PhysRevLett.103.150502}. 

Significant progress has been made in synthesizing quantum circuits for arbitrary pure-state preparation. By exploiting circuit symmetries, commutation relations, and structured decompositions, a variety of efficient gate-synthesis techniques have been developed \cite{araujo2023configurable,7515556}. Prominent examples include quantum multiplexors \cite{1629135}, universal gate decompositions \cite{PhysRevA.83.032302}, and isometry decompositions \cite{PhysRevA.93.032318}. In particular, the framework of uniformly controlled one-qubit gates \cite{PhysRevLett.93.130502} has led to systematic and near-optimal circuit constructions for preparing arbitrary $n$-qubit pure states \cite{PhysRevA.71.052330,mottonen2004transformatio}. Comprehensive benchmarks comparing different pure-state preparation methods can be found in \cite{benchmark}.

Further efficiency gains are possible when restricting attention to special classes of states exhibiting additional structure. Algorithms tailored to uniform \cite{9506863}, sparse \cite{Malvetti2021quantumcircuits,9586240,de2022double,PhysRevA.106.022617,10546633}, or probability-distribution states \cite{zoufal2019quantum} significantly outperform general-purpose approaches. Moreover, when approximate preparation is acceptable, several methods \cite{PhysRevResearch.4.023136,Creevey2023,10190145} provide favorable trade-offs between fidelity and circuit complexity. Such approximation-based strategies are particularly relevant in the noisy intermediate-scale quantum (NISQ) regime \cite{preskill2018quantum,brooks2019beyond}, where exact state preparation may be unnecessary or impractical.

\medskip

In contrast to pure states, the preparation of arbitrary mixed quantum states—described by density matrices—has received comparatively limited attention, with existing studies largely confined to fundamental concepts \cite{aharonov1998quantum,pinto2024preparing}. Mixed states naturally arise in realistic quantum systems due to environmental noise, decoherence, and incomplete knowledge of system dynamics \cite{alber2003quantum,liang2019quantum}. Their preparation is essential for modeling open quantum systems, analyzing entanglement and quantum correlations in noisy settings \cite{PhysRevA.67.022110,PhysRevA.75.042310}, and developing robust quantum algorithms and error-correction strategies \cite{PhysRevA.54.3824,PhysRevA.70.052309}. Furthermore, thermal states---central to studies of quantum thermalization \cite{PhysRevLett.108.080402,PhysRevE.103.062133} and quantum chemistry \cite{RevModPhys.92.015003,lanyon2010towards,cao2019quantum}---are inherently mixed, underscoring the importance of efficient mixed-state preparation.

This work focuses on systematic circuit synthesis methods for preparing mixed quantum states. When the target mixed state is specified as an ensemble of pure states,  we propose two approaches: one based on the mixing of pure states and the other based on purification. The former requires fewer qubits and is well suited for dynamic circuits \cite{PhysRevLett.127.100501,IBM:dynamic}, where mid-circuit measurement and qubit reuse are available, and for scenarios involving external quantum channels. The latter approach introduces additional ancilla qubits but yields a purified state of the target mixed state and can be further optimized using isometry decompositions or uniformly controlled one-qubit gates \cite{PhysRevA.93.032318,PhysRevLett.93.130502,PhysRevA.71.052330,mottonen2004transformatio}.

\medskip

It is worth emphasizing that, in the context of large-scale quantum computation, specifying an \emph{arbitrary} pure state or density matrix is neither practical nor meaningful: a generic $n$-qubit pure or mixed state requires $O(2^n)$ complex parameters or $O(4^n)$ real parameters, respectively, which quickly becomes infeasible beyond modest system sizes. The power of quantum computation therefore does not stem from the ability to prepare arbitrary states, but rather from efficiently preparing structured states relevant to specific computational tasks. Accordingly, extensive effort has been devoted to preparing highly or even extremely sparse pure states \cite{Malvetti2021quantumcircuits,9586240,de2022double,PhysRevA.106.022617,10546633}. By contrast, the preparation of sparse mixed quantum states remains largely unexplored and constitutes a central focus of this work.

Sparse mixed states may arise in different forms. In some settings, such as simulations of thermalization, a mixed state is naturally specified as an ensemble of a small number of sparse pure states \cite{PhysRevA.101.012328}. In this case, our proposed methods can be directly combined with existing sparse pure-state preparation techniques to obtain efficient circuit implementations.

In other applications, such as quantum machine learning, mixed states are often specified directly by sparse density matrices, for example those obtained via classical-shadow preprocessing or compressed-data reconstruction \cite{huang2020predicting,bermejo2024quantum}. To apply ensemble-based preparation methods in this setting, one must first convert the density-matrix representation into an ensemble of pure states. While a direct eigenvalue decomposition can achieve this conversion, it is computationally expensive. Instead, we introduce a novel strategy based on the \emph{Cholesky decomposition} \cite{trefethen1997numerical,gentle2012numerical,golub2012matrix,Matlab:chol}, which is substantially more efficient, especially for low-rank or sparse matrices.

Crucially, the Cholesky factorization matrix generically exhibits enhanced sparsity, even when the original density matrix is not sparse. This property allows existing sparse pure-state preparation methods \cite{Malvetti2021quantumcircuits,9586240,de2022double,PhysRevA.106.022617} to be leveraged to significantly reduce circuit complexity. When the target density matrix itself is sparse, the advantages are further amplified in both classical preprocessing and quantum circuit synthesis.

Moreover, by employing the \emph{incomplete Cholesky decomposition} with threshold dropping \cite{golub2012matrix,Matlab:ichol}, we obtain a flexible framework for approximating mixed states with high fidelity. Increasing the drop tolerance yields progressively sparser decompositions, leading to substantial gains in efficiency while maintaining controllable approximation errors. This provides a practical and tunable balance between accuracy and resource requirements.

\medskip

Beyond state preparation, we also investigate the closely related task of mixing or superposing quantum states. Given a set of arbitrary \emph{unknown} input states in a Hilbert space of dimension $d\geq2$, we ask whether a physical process can generate an output state that realizes a prescribed mixture or superposition with predefined weights or coefficients. While arbitrary unknown states can indeed be mixed according to predefined weights, we prove that, for pure states, no physical operation can produce a superposition of arbitrary unknown states with predefined coefficients unless all but one coefficient vanish. We refer to this result as the \emph{no-superposing theorem}.

This paper is organized as follows. In \secref{sec:mixture}, we study the problem of mixing arbitrary unknown quantum states and address the no-superposing theorem. In \secref{sec:arbitrary density matrix}, we present systematic schemes for preparing arbitrary mixed states, including approximation strategies. Performance benchmarks are provided in \secref{sec:benchmark}. We conclude in \secref{sec:summary}. Background material on pure-state preparation and uniformly controlled one-qubit gates is included in \appref{sec:arbitrary state} and \appref{sec:uniformly control}, respectively.

\section{Superposition and mixture of unknown quantum states}\label{sec:mixture}
Before undertaking the task of implementing quantum circuits capable of preparing arbitrary quantum states, let us first consider a different yet closely related challenge: provided with a certain number of arbitrary \emph{unknown} pure or mixed quantum states as input, can we transform them into a new output state as a superposition or mixture of the input states with predefined coefficients or weights? If this can be achieved, the associated strategies can be utilized for preparing arbitrary quantum states. 

\subsection{Superposition of pure states}\label{sec:superposition of states}
The task of superposing arbitrary unknown pure states with predefined coefficients turns out to be unattainable (by \emph{any} means, not merely by quantum circuits), except in a trivial exceptional case. We refer to this result as the \emph{no-superposing theorem}, which is formally stated and proven as follows.

\begin{theorem}[No-superposing theorem]\label{thm:no-superposing}
Given $\ell$ predefined coefficients $\alpha_i\in\mathbb{C}$ for $i=0,\dots,\ell-1$, and $\ell$ arbitrary \emph{unknown} pure quantum states $\ket{\psi_i}\in \mathcal{H}_i$, where the Hilbert spaces $\mathcal{H}_i$ are mutually isomorphic (i.e., $\mathcal{H}_i \cong \mathcal{H}_j \cong \mathcal{H}$ for all $i,j=0,\dots,\ell-1$) with dimension $d\ge 2$, it is impossible to devise an operation that generates a new state $\ket{\psi}\in\mathcal{H}$ such that
\begin{equation}\label{superposed state}
\ket{\psi} \propto \sum_{i=0}^{\ell-1}\alpha_i\ket{\psi_i},
\end{equation}
unless one is in the exceptional case in which all but one of the coefficients $\alpha_i$ vanish. Such an operation remains impossible even if it is required to succeed only probabilistically.
\end{theorem}

\begin{proof}
If such an operation were possible, the global phase of a quantum state would acquire physical significance. For instance, replacing $\ket{\psi_i}$ by $e^{i\theta}\ket{\psi_i}$ would yield a different resulting state $\ket{\psi}$ in \eqref{superposed state}. This would lead to a contradiction as shown below.

Since $d\ge2$, we can choose two orthogonal states and denote them by $\ket{0}$ and $\ket{1}$. Now consider the following scenario. Alice sends to Bob one of the two predefined states $\ket{\phi_a}=\ket{0}$ or $\ket{\phi_b}=e^{i\theta}\ket{0}$ and challenges Bob to determine which state has been sent.

Suppose that the operation described in the theorem were possible for the given coefficients $\alpha_i$. Bob could then construct such an operation that transforms any $\ell$ arbitrary input states $\ket{\psi_i}$ into a single output state $\ket{\psi}\propto\sum_{i=0}^{\ell-1}\alpha_i\ket{\psi_i}$. In particular, Bob prepares the inputs $\ket{\psi_i}=\ket{1}$ for $i=1,\dots,\ell-1$, while the input slot $\ket{\psi_0}$ receives the state sent by Alice.

If Alice sends $\ket{\phi_a}$ or $\ket{\phi_b}$, Bob's operation produces
\begin{equation}
\ket{\psi_a}\propto \alpha_0\ket{0}+\beta\ket{1},
\qquad
\ket{\psi_b}\propto \alpha_0 e^{i\theta}\ket{0}+\beta\ket{1},
\end{equation}
respectively, where $\beta=\sum_{i=1}^{\ell-1}\alpha_i$. Given the coefficients $\alpha_i$, one can always choose a value of $\theta$ such that $|\braket{\psi_a}{\psi_b}|<1$.

Since $|\braket{\psi_a}{\psi_b}|\neq1$, Bob can distinguish whether the resulting state is $\ket{\psi_a}$ or $\ket{\psi_b}$ with a nonzero success probability.\footnote{For instance, this can be achieved by unambiguous quantum state discrimination (UQSD) or minimum-error discrimination (MED).} Consequently, Bob could infer whether the state sent by Alice is $\ket{\phi_a}$ or $\ket{\phi_b}$ with a nonzero success rate equal to the success probability of realizing the superposing operation multiplied by the success probability of distinguishing $\ket{\psi_a}$ and $\ket{\psi_b}$. This contradicts the fact that $\ket{\phi_a}=\ket{0}$ and $\ket{\phi_b}=e^{i\theta}\ket{0}$ are physically indistinguishable.
Therefore, the operation described in the theorem is impossible, even probabilistically, except in the exceptional case.

In the exceptional case where all $\alpha_i$ vanish except one (say $\alpha_0$), the operation can be trivially implemented by retaining $\ket{\psi_0}$ while discarding $\ket{\psi_1},\dots,\ket{\psi_{\ell-1}}$.
\end{proof}

\subsection{Mixture of mixed states}
On the other hand, the task of mixing arbitrary unknown mixed states is possible. We first consider the challenge of mixing two unknown mixed state, formally stated as follows.
Given a predefined real number $0\leq p\leq1$, devise an operation that generates the new mixed state
\begin{equation}\label{eq:rho combine}
\rho = p \rho_0+ (1-p) \rho_1 \in \mathcal{L}(\mathcal{H}),
\end{equation}
when provided with any two arbitrary \emph{unknown} mixed quantum states $\rho_0\in \mathcal{L}(\mathcal{H}_0)$ and $\rho_1\in \mathcal{L}(\mathcal{H}_1)$, where $\mathcal{H}_0 \cong\mathcal{H}_1 \cong\mathcal{H}$.

In the case that both $\rho_0$ and $\rho_1$ are $n$-qubit mixed states carried by quantum wires, this operation can be achieved by implementing the circuit as depicted in \figref{fig:rho combine}.
First, we prepare the state $\sqrt{p}\ket{0}+\sqrt{1-p}\ket{1}$ for the ancilla qubit (the feasibility of this preparation will be discussed in \secref{sec:arbitrary state}). Then, we apply a sequence of $n$ controlled-SWAP (CSWAP, also known as Fredkin) gates that interchange $\rho_0$ and $\rho_1$ when the ancilla qubit is in $\ket{1}$ and do nothing when the ancilla qubit is $\ket{0}$.
By tracing out the second $n$-qubit wire and the ancilla wire, the circuit yields the state $\rho$ given by \eqref{eq:rho combine}.

\begin{figure}
\begin{quantikz}
\lstick{$\rho_0$} &\qwbundle{n}& \swap{1}& \qw  \rstick{$\rho$} \\
\lstick{$\rho_1$} &\qwbundle{n}&  \swap{1} &\trash{\text{trash}} \\
\lstick{$\sqrt{p}\ket{0}+\sqrt{1-p}\ket{1}$}&\qw & \phase{n}\vqw{-1} &\trash{\text{trash}}
\end{quantikz}
\caption{The quantum circuit transforming $\rho_0$ and $\rho_1$ into $\rho$ in \eqref{eq:rho combine}. Here, the vertical line adorned with one dot and two crosses represents a sequence of $n$ CSWAP gates applied to each pair of the corresponding qubits carrying $\rho_0$ and $\rho_1$.}
\label{fig:rho combine}
\end{figure}

The scheme of mixing two unknown mixed states can be straightforwardly extended to the case of many unknown mixed states. That is, given a set of $\ell$ predefined real numbers $0< p_i\leq 1$ for $i=0,\dots,\ell-1$ satisfying $\sum_{i=0}^{\ell-1} p_i=1$, produce the new mixed state
\begin{equation}\label{eq:density matrix via rho}
\rho = \sum_{i=0}^{\ell-1} p_i\, \rho_i \in \mathcal{L}(\mathcal{H}),
\end{equation}
when provided with any $\ell$ arbitrary \emph{unknown} mixed quantum states $\rho_i\in \mathcal{L}(\mathcal{H}_i)$, where $\mathcal{H}_i \cong\mathcal{H}_j \cong\mathcal{H}$ for all $i,j=0,\dots,\ell-1$.
The quantum circuit for mixing $\rho_i$ into $\rho$ is depicted in \figref{fig:circuit for rho via add}.
At each step in the circuit, $\rho_i$ is fed into the circuit and the ancilla is prepared in the state
\begin{equation}\label{eq: weight state}
\ket{\alpha_i} := \cos{\alpha_i} \ket{0} + \sin{\alpha_i} \ket{1},
\end{equation}
for $i = 1,\dots,\ell-1$,
where
\begin{equation}\label{alpha i}
\alpha_i = \arctan\left(\sqrt{\frac{p_i}{\sum_{j=0}^{i-1} p_j}}\right).
\end{equation}

This circuit requires $2n+1$ qubit registers, if the qubits that are discarded can be recycled, which is possible in the dynamic-circuit framework \cite{PhysRevLett.127.100501}, such as on transmon-based quantum computers in IBM Quantum \cite{IBM:dynamic}. On the other hand, if only a static circuit is feasible, such as in the case of a trapped-ion quantum computer, it requires $\ell(n+1)-1$ qubit registers. In both cases, the circuit requires $n(\ell-1)$ CSWAP gates and $\ell-1$ one-qubit rotations used to produce \eqref{eq: weight state} in the worst-case scenario.\footnote{A CSWAP gate can be decomposed into a Toffoli gate and two CNOT gates. More precisely, $\text{CSWAP}_{0,12}=\text{CNOT}_{2,1}\text{Toffoli}_{01,2}\text{CNOT}_{2,1}$, where the indices before the comma in the subscript represent the control qubit(s), and those after the comma represent the target qubit(s). A Toffoli gate can be further decomposed into 2 $H$ gates, 3 $T$ gates, 4 $T^\dag$ gates, 1 $S$ gate, and 6 CNOT gates (see Figure 4.9 in \cite{nielsen2010quantum}). Therefore, in terms of primitive gates, a CSWAP gate requires 8 CNOT gates and 10 one-qubit rotation gates in total.}

\begin{figure}
\begin{quantikz}
\lstick{$\rho_0$}
&\qwbundle{n}
&\swap{1}
&& \ \ldots\ &&&&
&\swap{1}
&\rstick{$\rho$}\\
\lstick{$\rho_1$}
&\qwbundle{n}
& \targX{}
&\trash{\text{trash}}
&\setwiretype{n}\ \ldots\ &&
& \lstick{$\rho_{\ell-1}$}
&\setwiretype{q}\qwbundle{n}
& \targX{}&\trash{\text{trash}}\\
\lstick{$ \ket{\alpha_1} $}&
& \ctrl{-1}
&\trash{\text{trash}}
&\setwiretype{n}\ \ldots\ &&
&\lstick{$ \ket{\alpha_{\ell-1}} $}&\setwiretype{q}
&\ctrl{-1}&\trash{\text{trash}}
\end{quantikz}
\caption{The quantum circuit transforming a set of mixed states $\rho_i$ into $\rho$ in \eqref{eq:density matrix via state}.}
\label{fig:circuit for rho via add}
\end{figure}

As a pure state $\ket{\psi}$ can be considered a special case of a mixed state, i.e.\ $\rho=\ket{\psi}\bra{\psi}$, it should be noted that, when provided with $\ell$ arbitrary unknown states $\ket{\psi_i}$ for $i=0,\dots,\ell-1$, we can devise a quantum operation to generate the mixed state $\rho=\sum_{i=0}^{\ell-1}p_i\ket{\psi_i}\bra{\psi_i}$, given any predefined real numbers $p_i$ satisfying $0\leq p_i\leq1$ and $\sum_{i=0}^{\ell-1}p_i=1$, but it is impossible to devise a mechanism to generate the pure state proportional to $\sum_{i=0}^{\ell-1}\alpha_i\ket{\psi_i}$, given the predefined coefficients $\alpha_i\in\mathbb{C}$, unless all but one of $\alpha_i$ vanish.

\section{Preparation of arbitrary mixed states}\label{sec:arbitrary density matrix}
We now turn to our main focus: quantum circuit synthesis for the preparation of arbitrary mixed states (i.e., density matrices). A density matrix may be specified in two equivalent ways: either as an ensemble of pure states or directly through its matrix elements. We treat these two representations separately, as they lead to distinct circuit synthesis strategies. As foundational subroutines, we review the preparation of arbitrary pure states in \appref{sec:arbitrary state} and the construction of uniformly controlled one-qubit gates in \appref{sec:uniformly control}.

\subsection{Density matrix specified as an ensemble of pure states}\label{sec:density matrix as ensemble}
A density matrix $\rho\in\mathcal{L}(\mathcal{H})$ can be specified as
\begin{equation}\label{eq:density matrix via state}
\rho = \sum_{i=0}^{\ell-1} p_i \ket{\psi_i}\bra{\psi_i},
\end{equation}
given a set of $\ell$ real numbers $0< p_i\leq 1$ satisfying $\sum_{i=0}^{\ell-1} p_i=1$ and a set of $\ell$ pure states $\ket{\psi_i}\in\mathcal{H}$ for $i=0,\dots,\ell-1$.
We may call $\{p_i, \ket{\psi_i}\}$ an \emph{ensemble of pure states}. Accordingly, the mixed state $\rho$ can be understood as being in the state $\ket{\psi_i}$ with the respective probability $p_i$.
However, the same $\rho$ admits different ensembles of pure states, a property known as \emph{unitary freedom} (see Theorem 2.6 in \cite{nielsen2010quantum} for more details).
Particularly, $\rho$ can be diagonalized into
\begin{equation}\label{eq:density matrix via eigenstates}
\rho = \sum_{i=0}^{\text{rank}(\rho)-1} \lambda_i\ket{\lambda_i}\bra{\lambda_i},
\end{equation}
where the nonzero eigenvalues $0<\lambda_i\leq1$ satisfy $\sum_i\lambda_i=1$, and the eigenstates $\ket{\lambda_i}$ are orthonormal. The mixed state $\rho$ can then be understood in terms of the ensemble of pure states $\{\lambda_i,\ket{\lambda_i}\}$.

Given a density matrix $\rho$ as specified in \eqref{eq:density matrix via state}, we can perform preprocessing to recast it into \eqref{eq:density matrix via eigenstates} before considering the implementation of the quantum circuit that produces $\rho$. However, whether doing so is advantageous or not depends on whether the preprocessing is time-consuming, whether the rank of $\rho$ is smaller than $\ell$, and whether the states $\ket{\lambda_i}$ are easier to prepare than the states $\ket{\psi_i}$.
As \eqref{eq:density matrix via eigenstates} can be viewed as a special case of \eqref{eq:density matrix via state}, in the following, we adhere to \eqref{eq:density matrix via state} and provide two methods for preparing $\rho$. Which of the two methods is more efficient depends on the hardware.

\subsubsection{Method via a mixture of pure states}
Obviously, \eqref{eq:density matrix via state} can be understood as \eqref{eq:density matrix via rho} with $\rho_i=\ket{\psi_i}\bra{\psi_i}$ for $i=0,\dots,\ell-1$. Therefore, the circuit in \figref{fig:circuit for rho via add} can be used to prepare $\rho$, if each $\rho_i$ is provided with the pure state $\ket{\psi_i}$.
The states $\ket{\psi_i}$, for $i=0,\dots,\ell-1$, can be prepared using a variety of existing pure-state preparation methods. Among these, one may optionally choose methods tailored to the specific structural features of $\ket{\psi_i}$ to achieve improved efficiency.

The circuit in \figref{fig:circuit for rho via add} requires $2n+1$ qubit registers in the dynamic circuit framework and $\ell(n+1)-1$ qubit registers in the static circuit framework.
In the worst-case scenario, the circuit requires $\ell(2^{n+1}-2n-2)$ CNOT gates, $\ell(2^{n+1}-2)+(\ell-1)=\ell(2^{n+1}-1)-1$ one-qubit rotation gates, and $n(\ell-1)$ CSWAP gates in total, including the components for preparing the states $\rho_i=\ket{\psi_i}\bra{\psi_i}$ and $\ket{\alpha_i}$, assuming that the arbitrary pure-state preparation circuit described in \secref{sec:arbitrary state} is used to prepare each $\ket{\psi_i}$.

\subsubsection{Method via purification with uniformly controlled rotations}
According to the \emph{purification theorem} \cite{nielsen2010quantum}, any arbitrary state $\rho_A$ in a finite-dimensional Hilbert space $\mathcal{H}_A$ can always be \emph{purified} into a pure state $\ket{\Psi_{AB}}$ in an enlarged Hilbert space $\mathcal{H}_{AR}\cong{\mathcal H}_A\otimes\mathcal{H}_R$ such that $\rho_A$ is a partial trace of $\ket{\Psi_{AR}}$, i.e., $\rho_A=\Tr_R \ket{\Psi_{AR}}\bra{\Psi_{AR}}$.
Particularly, with the inclusion of $m =\lceil\log_2 \ell\rceil$ ancilla qubits, the state given by \eqref{eq:density matrix via state} can be purified into
\begin{equation}\label{rho purification}
\ket{\Psi}^{(n+m)} = \sum_{i=0}^{\ell-1} \sqrt{p_i} \ket{\psi_i}^{(n)}\otimes\ket{i}^{(m)},
\end{equation}
where $\ket{i}^{(m)}$ are the computational basis states of the ancilla qubits.
The state $\rho$ then can be obtained simply by ignoring the ancilla qubits.
Let $U[\ket{\psi}^{(k)}]$ denote a unitary circuit that maps the state $\ket{0}^{\otimes k}$ to the desired arbitrary $k$-qubit state $\ket{\psi}^{(k)}$; the preparation of $\rho$ is achieved by synthesizing a unitary gate $U[\ket{\Psi}^{(n+m)}]$.

Instead of synthesizing a $U[\ket{\Psi}^{(n+m)}]$ gate directly, we propose a slightly more efficient scheme to obtain $\ket{\Psi}^{(n+m)}$ as follows.
First, we prepare the state
\begin{equation}\label{p state}
\ket{P}^{(m)} := \sum_{i=0}^{\ell-1}  \sqrt{p_i}\ket{i}^{(m)}
\end{equation}
for the ancilla qubits by applying $U[\ket{P}^{(m)}]$ on $\ket{0}^{(m)}$.
Next, with the ancilla qubits serving as the control nodes, we sequentially apply controlled-$U[\ket{\psi_i}]$ gates conditioned on the corresponding control node configurations as illustrated in \figref{fig:circuit for rho via cU}.
Finally, simply ignoring the ancilla qubits, we obtain the state given in \eqref{eq:density matrix via state}.
In comparison to the method via a mixture of pure states, the method via purification offers an additional bonus of providing a purified state, which could be useful in various situations.

As reviewed in \appref{sec:uniformly control}, each $U[\ket{\psi_i}]$ gate can be implemented in a series of $k$-fold \emph{uniformly controlled rotations} $F^{k}[R_y]$ and $F^{k}[R_z]$ as depicted in \figref{fig:circuit for n-qubit state}. Consequently, each $m$-fold controlled-$U[\ket{\psi_i}]$ gate appearing in \figref{fig:circuit for rho via cU} can be implemented in the same way, except that the uniformly controlled configurations in \figref{fig:circuit for n-qubit state} are all enlarged to include the corresponding control node configuration of the $m$ ancilla qubits.
Since controlled-gates conditioned on distinct control node configurations commute with one another, by rearranging the sequence of the controlled-$R_y$ and controlled-$R_z$ gates, the circuit in \figref{fig:circuit for rho via cU} collectively can be rendered into a layout similar to \figref{fig:circuit for n-qubit state} but with the uniformly controlled rotations $F^{k}[R_y]$ and $F^{k}[R_z]$ for $k=m,\dots,m+n-1$ instead of $k=0,\dots,n-1$.

In the worst-case scenario, the circuit of \figref{fig:circuit for rho via cU} requires $\sum_{k=m}^{n+m-1}(2\times2^k-2)=2^m(2^{n+1}-2)-2n$ CNOT gates and $\sum_{k=m}^{n+m-1}(2\times2^k)=2^m(2^{n+1}-2)$ one-qubit rotations.
Additionally, to prepare the state $\ket{P}^{(m)}$ in \eqref{p state}, it requires $\sum_{k=0}^{m-1}(2^k)=2^m-1$ CNOT gates and $\sum_{k=0}^{m-1}(2^k)=2^m-1$ one-qubit $R_y$ gates in the worst case.\footnote{As the coefficients in \eqref{p state} are all chosen to be real, no $R_z$ gates are needed.}
To sum up, in the worst case, the circuit for preparing an arbitrary $n$-qubit state $\rho$ requires $2^m(2^{n+1}-1)-2n-1\sim\ell(2^{n+1}-1)-2n-1$ CNOT gates and $2^m(2^{n+1}-1)-1\sim\ell(2^{n+1}-1)-1$ one-qubit rotation gates in total.\footnote{This is slightly more efficient than directly implementing the $U[\ket{\Psi}^{(n+m)}]$ gate, which requires $2^{n+m+1}-2(m+n)-2$ CNOT gates and $2^{n+m+1}-2$ one-qubit rotation gates using the method in \secref{sec:arbitrary state}.}


\begin{figure}
\begin{quantikz}
\lstick{$\ket{0}^{(n)}$} &\qwbundle{n}& \gate{U[\ket{\psi_0}]} & \gate{U[\ket{\psi_1}]}&\qw\ \ldots\ &\gate{U[\ket{\psi_{\ell-1}}]}&\qw &\qw  \rstick{$\rho$} \\
\lstick[4]{$\ket{P}^{(m)}$}&\qw & \octrl{-1}& \ctrl{-1}&\qw\ \ldots\ &\ctrl{-1} &\qw&\qw\rstick[4]{trash}\\
&\qw &\octrl{-1}\vqw{1}&\octrl{-1}\vqw{1}&\qw\ \ldots\ &\ctrl{-1}\vqw{1} &\qw&\qw\\
\wave&&&&&&&\\
&\qw &\octrl{-1}&\octrl{-1}&\qw\ \ldots\ &\ctrl{-1} \slice{$\;\;\;\ket{\Psi}^{(n+m)}$}&\qw&\qw
\end{quantikz}
\caption{The quantum circuit producing the mixed state $\rho$ in \eqref{eq:density matrix via state} and the corresponding purified state $\ket{\Psi}^{(n+m)}$ in \eqref{rho purification}.}
\label{fig:circuit for rho via cU}
\end{figure}

\subsubsection{Method via purification with isometries}\label{sec:purification with isometries}
A linear map $f : X \to Y$ between two complex inner product spaces (e.g., Hilbert spaces) is called an \emph{isometry} (or norm-preserving map) if, for any $\ket{\psi_1}, \ket{\psi_2} \in X$,
\begin{equation}
\langle \psi_1 | \psi_2 \rangle_X = \langle \psi'_1 | \psi'_2 \rangle_Y,
\end{equation}
where $\ket{\psi'_1} = f\ket{\psi_1}$ and $\ket{\psi'_2} = f\ket{\psi_2}$.

Efficient quantum circuit synthesis for arbitrary isometries mapping $m$ qubits to $n$ qubits has been extensively studied~\cite{PhysRevA.93.032318}. In addition, circuit constructions specifically tailored to \emph{sparse} isometries have also been proposed~\cite{Malvetti2021quantumcircuits}. In particular, given a set of orthonormal $m$-qubit states $\{\ket{\psi_i}^{(m)}\}$ and a corresponding set of orthonormal $n$-qubit states $\{\ket{\psi'_i}^{(n)}\}$ for $i=0,1,\dots,\ell-1$, one can efficiently synthesize a quantum circuit implementing the isometric mapping
\begin{equation}\label{psi to psiprime}
\ket{\psi_i}^{(m)} \mapsto \ket{\psi'_i}^{(n)}, \qquad i=0,1,\dots,\ell-1.
\end{equation}

As a special case with $\ell = 1$, an isometry that maps $\ket{0}^{(m)}$ to an arbitrary $n$-qubit state $\ket{\psi'}^{(n)}$ can be used for pure-state preparation. This strategy achieves circuit efficiency comparable to methods based on uniformly controlled one-qubit gates~\cite{PhysRevA.93.032318,PhysRevLett.93.130502,PhysRevA.71.052330,mottonen2004transformatio}. For a mixed state $\rho$ specified in the form of \eqref{eq:density matrix via state}, one may therefore construct an isometry circuit that maps $\ket{0}$ to the purified state in \eqref{rho purification} as a direct method for preparing $\rho$.

When a mixed state $\rho$ is specified as an ensemble of \emph{orthonormal} pure states---namely in the form of \eqref{eq:density matrix via eigenstates}---its preparation via purification can potentially be carried out more efficiently using the method proposed in~\cite{pinto2024preparing}. Furthermore, if these states are sparse, circuits for sparse isometries~\cite{Malvetti2021quantumcircuits} can be employed to achieve a substantial reduction in circuit complexity. In what follows, we briefly outline this synthesis scheme and analyze the associated CNOT gate count; a detailed discussion can be found in~\cite{pinto2024preparing}.

The overall quantum circuit of the method in~\cite{pinto2024preparing} is illustrated in \figref{fig:circuit for rho via isometry} and can be divided into three steps. Let $m = \lceil \log_2 \ell \rceil$, and let $\ket{P_\lambda}^{(m)}$ denote the state defined in \eqref{p state} with $p_i = \lambda_i$.

First, prepare the pure state $\ket{P_\lambda}^{(m)}$.
Second, apply CNOT gates bitwise between this register and $m$ ancilla qubits initialized in the state $\ket{0}^{(m)}$, yielding
\begin{equation}
{|\tilde{P}_\lambda\rangle}^{(2m)} = \sum_{i=0}^{\ell-1} \sqrt{\lambda_i}\ket{i}^{(m)} \otimes \ket{i}_a^{(m)}.
\end{equation}
Third, with the inclusion of additional $n-m$ qubits initialized to $\ket{0}^{(n-m)}$, apply an isometry that implements the mapping
\begin{equation}
\ket{i}^{(m)} \mapsto \ket{\lambda_i}^{(n)}, \qquad i=0,1,\dots,\ell-1,
\end{equation}
as a specific instance of \eqref{psi to psiprime}. This results in the purified state
\begin{equation}\label{eq:rho purification oth}
\ket{\Psi}^{(n+m)} = \sum_{i=0}^{\ell-1} \sqrt{\lambda_i}\ket{\lambda_i}^{(n)} \otimes \ket{i}_a^{(m)}.
\end{equation}
By tracing out (i.e., ignoring) the ancilla register, one obtains the target mixed state $\rho$. If the state $\ket{P_\lambda}^{(m)}$ exhibits additional structure, state-preparation schemes tailored to these features can be employed to further reduce circuit complexity.

\begin{figure}
\centering
\begin{quantikz}
\lstick{$\ket{0}^{(m)}$}&\qwbundle{m}& \gate{U[\ket{P_\lambda}^{(m)}]}&
\ctrl{2} &\gate[2]{\text{Isometry}}&\qw \qw&  \rstick[2]{$\rho$} \\
\lstick{$\ket{0}^{(n-m)}$}&\qwbundle{n-m}& \qw& \qw&  \qw& \qw&  \\
\lstick{$\ket{0}^{(m)}_a$}&\qwbundle{m}& \qw& \targ{}& \qw& \qw\slice{$\;\;\;\;\;\ket{\Psi}^{(n+m)}$}&  \trash{\text{trash}}
\end{quantikz}
\caption{Quantum circuit for preparing the mixed state in \eqref{eq:density matrix via eigenstates} via purification, yielding the purified state in \eqref{eq:rho purification oth}. The CNOT gate shown represents a sequence of $m$ CNOT gates applied bitwise.}
\label{fig:circuit for rho via isometry}
\end{figure}

\begin{table}
\centering
\renewcommand{\arraystretch}{1.3}
\begin{tabular}{l|c}
\hline\hline
Method & CNOT count upper bound \\
\hline\hline
A: Isometry mapping $\ket{0}$ to \eqref{rho purification} & $\frac{23}{24} 2^{m+n} - 2^{(m+n)/2 +1} +  O((m+n)^2)$  \\
B: \figref{fig:circuit for rho via isometry} with CCD isometry & $2^{m+n} + O(n^2) 2^{m}$  \\
C: \figref{fig:circuit for rho via isometry} with CSD isometry & $\frac{23}{144}(4^m + 4^n+1) + 2^m + m -1$ \\
\hline
D: Sparse isometry mapping $\ket{0}$ to \eqref{rho purification} &  $(n+m+16s-9)\,\mathrm{NNZ}+ \frac{23}{24} 2^s$  \\
E: \figref{fig:circuit for rho via isometry} with sparse isometry & $(7n+4)\,\mathrm{NNZ} + (14n+24m-\frac{481}{24}) 2^m -2^{m/2+1}+m +\frac{5}{3}  $ \\
\hline\hline
\end{tabular}
\caption{Theoretical upper bounds on the CNOT gate count required to produce the purified state in \eqref{eq:rho purification oth} using different methods. Here, $s := \lceil \log_2(\mathrm{NNZ}) \rceil$. Methods A, B, and C employ circuit synthesis methods for generic isometries, whereas Methods D and E employ synthesis methods tailored for sparse isometries.
}
\label{tab:gate_count}
\end{table}

The total CNOT gate count required by this construction scheme is summarized in Table~\ref{tab:gate_count}. The count includes contributions from preparing $\ket{P_\lambda}^{(m)}$, performing the bitwise CNOT operations, and implementing the isometry circuit. In our analysis we do not exploit specialized preparation methods for structured states $\ket{P_\lambda}^{(m)}$; instead, we assume $\ket{P_\lambda}^{(m)}$ to be arbitrary, which requires $2^m-1$ CNOT gates for its preparation.

The circuit complexity of isometries depends on the chosen matrix decomposition method, as discussed in~\cite{PhysRevA.93.032318}. Here we adopt the column-by-column decomposition (CCD) and the cosine--sine decomposition (CSD) for implementing isometries from $m$ to $n$ qubits with $1 \le m \le n-2$. When the target density matrix $\rho$ is extremely sparse, circuit synthesis methods tailored for sparse isometries can be employed to further reduce the circuit complexity. In this case, the complexity primarily depends on the number of nonzero matrix elements (NNZ) of $\rho$, and additional ancilla qubits may be required (see~\cite{Malvetti2021quantumcircuits} for details).

When the qubit number $n$ becomes large, the direct isometry-based purification methods (i.e., Methods~A and~D) generally yield more efficient circuits---typically in terms of the CNOT gate count---than the methods illustrated in \figref{fig:circuit for rho via isometry} (i.e., Methods~B, C, and~E). The latter approaches, however, may offer advantages when additional structure in $\ket{P_\lambda}^{(m)}$ can be exploited, or when $n$ is of moderate size.\footnote{For instance, compare the upper bounds of Methods~A, B, and~C shown in the upper-left panels of \figref{fig:cx count q10} and \figref{fig:cx count q12}.}


\subsection{Density matrix specified by matrix elements}
In many applications, the density matrix $\rho$ is not given in the form of \eqref{eq:density matrix via state}, but specified by its matrix elements, i.e.,
\begin{equation}\label{eq:density matrix component form}
\rho = \sum_{a,b=0}^{2^n-1}\rho_{ab}\ket{a}\bra{b},
\end{equation}
where $\ket{a}$ are the $n$-qubit computational basis states, and
\begin{equation}
\rho_{ab} = \rho^*_{ba} = \bra{a}\rho\ket{b}.
\end{equation}
Given $\rho_{ab}$, can we design a circuit that efficiently prepares the state $\rho$?\footnote{As noted in \secref{sec:introduction}, element-wise specification of an arbitrary density matrix quickly becomes infeasible beyond modest system sizes. In contrast, specifying a highly (or even extremely) sparse density matrix with only a small number of nonzero elements remains meaningful and practically relevant in many applications.}

An obvious strategy is to perform preprocessing to recast \eqref{eq:density matrix component form} into the diagonal form of \eqref{eq:density matrix via eigenstates} by solving the eigenvalue problem, followed by the application of the methods presented in \secref{sec:density matrix as ensemble}. However, this approach comes with two potential drawbacks. First, solving the eigenvalue problem can be time-consuming. Second, the eigenstates $\ket{\lambda_i}$ may be difficult to deal with when they are incorporated into the circuit. In the following, we present an alternative approach that circumvents the need to solve the eigenvalue problem and is more advantageous in both preprocessing and circuit implementation.

Suppose that $\rho$ can be efficiently recast into the form of \eqref{eq:density matrix via state}. By expanding each $\ket{\psi_i}$ as
\begin{equation}\label{psi expansion}
\ket{\psi_i} = \sum_{a=0}^{2^n-1}\alpha_{ai}\ket{a},
\end{equation}
where $\alpha_{ai}\in\mathbb{C}$ and $\sum_{a=0}^{2^n-1}|\alpha_{ai}|^2=1$, it leads to
\begin{eqnarray}
\rho
&=& \sum_{i=0}^{\ell-1}\sum_{a,b=0}^{2^n-1} p_i\alpha_{ai}\alpha_{bi}^* \ket{a}\bra{b} \nonumber\\
&=:& \sum_{i=0}^{\ell-1}\sum_{a,b=0}^{2^n-1} \tilde{\alpha}_{ai}\tilde{\alpha}_{bi}^* \ket{a}\bra{b},
\end{eqnarray}
where we have defined $\tilde{\alpha}_{ai}:=\sqrt{p_i}\,\alpha_{ai}$, and the values of $\ell$, $p_i$, and $\alpha_{ai}$ are to be determined from the given $\rho_{ab}$.
Equivalently, we have
\begin{equation}\label{rho factorization}
\rho = A A^\dag,
\end{equation}
where $\rho$ is viewed as a $2^n\times2^n$ matrix and $A$ is a $2^n\times\ell$ matrix with the matrix elements
\begin{equation}\label{Aai}
A_{ai}=\tilde{\alpha}_{ai}\equiv\sqrt{p_i}\,\alpha_{ai}.
\end{equation}
Since it is obvious that $A^\dag A$ is positive semi-definite and $\Tr A^\dag A=\Tr AA^\dag=\Tr \rho=1$, we have
\begin{equation}
0\leq(A^\dag A)_{ii} = \sum_{a=0}^{2^n-1} \tilde{\alpha}_{ai}\tilde{\alpha}_{ai}^* \leq 1
\end{equation}
and $\sum_i(A^\dag A)_{ii}=1$.
Consequently, letting
\begin{equation}\label{pi from A}
p_i=(A^\dag A)_{ii}
\end{equation}
leads to $0\leq p_i\leq1$ with $\sum_{i=0}^{\ell-1}p_i=1$ and $\sum_{a=0}^{2^n-1} |\alpha_{ai}|^2=1$ if $p_i>0$. The real numbers $p_i$ can be viewed as disjoint probabilities, and the complex numbers $\alpha_{ai}$ for $p_i>0$ can be viewed as normalized coefficients of an $n$-qubit state $\ket{\psi_i}$.

Therefore, if we can find a $2^n\times\ell$ matrix $A$ satisfying \eqref{rho factorization}, then the state $\rho$ can be produced by the methods discussed in \secref{sec:density matrix as ensemble}. The method via a mixture of pure states implements the circuit of \figref{fig:circuit for rho via add} with the following prescription: the real numbers $p_i>0$ given by \eqref{pi from A} are used to specify the states $\ket{\alpha_i}$ via \eqref{eq: weight state} and \eqref{alpha i}, and $\rho_i=\ket{\psi_i}\bra{\psi_i}$ are prepared as the states $\ket{\psi_i}$ given by \eqref{psi expansion} with the coefficients $\alpha_{ai}=A_{ai}/\sqrt{p_i}$.\footnote{If $p_i=0$, we have $\sum_{a=0}^{2^n-1} |\tilde{\alpha}_{ai}|^2=0$, implying $\tilde{\alpha}_{ai}=0$ for all $a=0,\dots,2^n-1$. The whole column of $\tilde{\alpha}_{ai}$ can be removed from $A$ to still satisfy \eqref{rho factorization}. Thus, the case of $p_i=0$ can be disregarded.}
On the other hand, the method via purification either implements the circuit of \figref{fig:circuit for rho via cU} or \figref{fig:circuit for rho via isometry} or employs other pure state preparation methods to produce the purified state $\ket{\Psi}^{(n+m)}$ given in \eqref{rho purification} with the number of ancilla qubits chosen to be $m=\lceil\log_2\ell\rceil$.
The only remaining challenge is to find the matrix $A$, which can be efficiently solved by the \emph{Cholesky decomposition} \cite{trefethen1997numerical, gentle2012numerical, golub2012matrix, Matlab:chol}.

\subsubsection{Cholesky decomposition}
In the Cholesky decomposition, if $M$ is a $d\times d$ Hermitian positive semi-definite matrix, then it admits a factorization
\begin{equation}
M = LL^\dag,
\end{equation}
where $L$ is a $d\times d$ lower triangular matrix.
Performing the Cholesky decomposition upon $\rho$ to obtain $L$ and then removing the columns that contain all zeros from $L$, we obtain a $2^n\times\ell$ matrix, which obviously is a solution of $A$ in \eqref{rho factorization}.
The factorization is unique if $M$ is positive definite, but need not be if $M$ is only positive semi-definite.
However, if the rank of $M$ is $r$, there is a unique factorization where $L$ is a $d\times d$ lower triangular matrix with $r$ positive diagonal elements and $d-r$ columns containing all zeros \cite{gentle2012numerical}.
This unique factorization leads to a solution of $A$ as a $2^n\times r$ matrix.\footnote{\label{foot:rank}The unique factorization gives the minimum possible matrix size of $A$ (i.e., $\ell=r$). However, the solutions with $\ell>r$ are still considered as they may exhibit higher sparsity, whose merit will be discussed shortly.}

Several algorithms exist for solving the Cholesky decomposition, each with its own strengths. Commonly used algorithms are the Cholesky algorithm, the Cholesky--Banachiewicz algorithm, and the Cholesky--Crout algorithm, all of which have a time complexity of $O(d^3)$, requiring $\sim d^3/3$ floating-point operations \cite{trefethen1997numerical}. The computational efficiency can depend on various factors, including the size and condition of the matrix, as well as other specific implementation details. 
If the rank of the matrix is sufficiently low, it can lead to significant computational efficiency. The pivoted Cholesky decomposition \cite{higham1990analysis,golub2012matrix} and the rank-revealing Cholesky decomposition \cite{gu2004strong} can be used to take advantage of the rank deficiency. Both rank-aware algorithms still have the time complexity of $O(d^3)$, but they can be significantly more efficient in practice for low-rank matrices.
Generally, the Cholesky decomposition is considered to be efficient, especially when compared to other matrix factorization methods.

Alternatively, we can perform the preprocessing to render \eqref{eq:density matrix component form} into the diagonal form of \eqref{eq:density matrix via eigenstates} by solving the eigenvalue problem. A commonly used algorithm for finding eigenvalues and eigenvectors is the QR algorithm \cite{trefethen1997numerical,gentle2012numerical}, which utilizes the QR decomposition and is efficient for dense matrices with a time complexity of $O(d^3)$, requiring $\sim 6d^3$ floating-point operations \cite{trefethen1997numerical}.
In comparison, solving the Cholesky decomposition in general is more efficient than solving the eigenvalue problem by a factor of $\sim18$. In the case where the qubit number $n$ is large enough, the matrix size $d=2^n$ can be huge, and the difference between the time consumption of $\sim d^3/3$ and $\sim 6d^3$ floating-point operations can be significant.

For solving the eigenvalue problems of sparse matrices, specialized iterative methods such as the Lanczos algorithm \cite{golub2012matrix,ojalvo1988origins}  may be more efficient in practice than the QR algorithm.
On the other hand, for solving the Cholesky decomposition of sparse matrices, one can significantly boost computational efficiency by applying the techniques of reordering strategies to obtain a sparse Cholesky factorization \cite{Matlab:chol,davis:direct,toledo:note}.
Therefore, regardless of whether $\rho$ is sparse or not, the preprocessing via solving the Cholesky decomposition is preferable to that via solving the eigenvalue problem.

It is important to note that, in general, the matrix $A$ is fairly sparse, as it is obtained from a lower triangular matrix, even if the original matrix $M$ is not sparse itself.\footnote{\label{foot:zeros}In the worst case, there are still at least $1+2+\dots+(\ell-1)=\ell(\ell-1)/2$ zeros in $A$.}
When the original matrix is sparse, reordering strategies can be employed to minimize the creation of ``fill-ins'' (i.e., the non-zero elements in the factorized matrices that are zero in the original matrix) and maintain sparsity in the factorized form as much as possible \cite{davis:direct,toledo:note}. These strategies aim to reorder the rows and columns of the matrix in a way that preserves its sparsity structure, reducing the number of non-zero entries introduced during the factorization.
By contrast, the coefficients $\beta_{ai}$ for the eigenstates $\ket{\lambda_i} =\sum_{a=0}^{2^n-1}\beta_{ai}\ket{a}$ do not always exhibit sparsity, even if the original matrix is sparse. The sparsity of a matrix does not necessarily imply sparsity in its eigenvectors, and vice versa.

The sparsity of the matrix $A$ offers a significant advantage in circuit implementation. With many of the coefficients $\alpha_{ai}$ vanishing, the corresponding states $\ket{\psi_i}$ are mostly sparse in the computational basis $\{\ket{a}\}$. Employing the algorithms specifically designed for preparing sparse states, such as those proposed in \cite{Malvetti2021quantumcircuits, 9586240, de2022double, PhysRevA.106.022617}, to prepare the pure states $\rho_i=\ket{\psi_i}\bra{\psi_i}$ in \figref{fig:circuit for rho via add} will markedly enhance circuit efficiency. Alternatively, as the purified state $\ket{\Psi}^{(n+m)}$ in \eqref{rho purification} as a whole is also sparse in the computational basis $\{\ket{a}^{(n)}\otimes\ket{i}^{(m)}\}$, these algorithms can also be applied to implement a highly efficient circuit for directly producing the state $\ket{\Psi}^{(n+m)}$.\footnote{If the rank $r$ of the given density matrix $\rho$ is sufficiently low, it can be preferable to apply the unique factorization to yield $A$ with the minimum possible matrix size $2^n\times r$. On the other hand, if the matrix $\rho$ is sufficiently sparse, it becomes preferable to apply the reordering strategies \cite{davis:direct,toledo:note} to yield $A$ with a larger matrix size but higher sparsity. (Recall \ftref{foot:rank}.)}

More importantly, as noted previously, in many relevant applications beyond modest system sizes, the practical demand is to prepare highly or even extremely sparse mixed states. In such scenarios, solving the eigenvalue problem---even for sparse matrices---is unfavorable and even inadequate, since the resulting eigenvectors need not preserve sparsity and may substantially increase the effective complexity of the state representation. By contrast, the Cholesky-based approach directly exploits, and in some cases further enhances, the sparsity structure of the density matrix, thereby leading to more efficient classical preprocessing and quantum circuit implementation.

\subsubsection{Incomplete Cholesky decomposition}\label{sec:icho}
In many applications, it is sufficient to prepare only an approximate state $\rho'$ that is close enough to $\rho$, instead of $\rho$ itself as originally specified.
The \emph{incomplete Cholesky decomposition} \cite{golub2012matrix, Matlab:ichol} can be performed on $\rho$ to obtain such an approximation.

For a $d\times d$ Hermitian positive semi-definite matrix $M$, the incomplete Cholesky decomposition gives the factorization
\begin{equation}
M' := L'L^{\prime\dag},
\end{equation}
where $L'$ is a $d\times d$ lower triangular matrix subject to a specified option for approximation, and $M'$ is ``close'' to $M$ in this specified sense.
Particularly, if the \emph{threshold dropping} option \cite{Matlab:ichol} is employed with a specified nonnegative drop tolerance $\epsilon$, any off-diagonal elements smaller in magnitude than $\epsilon$ are dropped from $L'$, and the resulting $M'$ is close to $M$ in the sense that
\begin{equation}\label{ichol approx}
\frac{\|M-M'\|_F}{\|M\|_F} \sim O(\epsilon),
\end{equation}
where $\|\cdot\|_F$ is the Frobenius norm.
As the drop tolerance $\epsilon$ increases, the accuracy of the approximation decreases, but the sparsity of $L'$ increases.

The incomplete Cholesky decomposition, prescribed with a considerably large drop tolerance, is extremely cheap in computational cost and yields a highly sparse factor matrix $L'$.
When dealing with a density matrix $M=\rho$, a highly sparse $2^n\times \ell$ matrix $A'$ can be obtained from $L'$ in the same manner that $A$ is obtained from $L$.
Consequently, we obtain a good approximation $\rho'$ to $\rho$ as
\begin{equation}
\rho' = \sum_{i=0}^{\ell-1} p_i\ket{\psi'_i}\bra{\psi'_i},
\end{equation}
where\footnote{Note that $\Tr A^{\prime\dag}A'=\Tr A'A^{\prime\dag} = \Tr M'$ is close to but slightly different from $\Tr M\equiv\Tr\rho=1$. Thus, we have to normalize $p'_i$ to $p_i$ to satisfy $\sum_{i=0}^{\ell-1}p_i=1$. Because of this, also note that $\rho'\neq M'$ but $\rho'=M'/\Tr M'\approx M'$.}
\begin{equation}
p'_i=(A^{\prime\dag}A')_{ii},
\quad
p_i = \frac{p'_i}{\sum_{i=0}^\ell p'_i},
\end{equation}
\begin{equation}
\ket{\psi'_i} = \sum_{a=0}^{2^n-1} \alpha'_{ai}\ket{a},
\end{equation}
and
\begin{equation}
A'_{ai}= \sqrt{p'_i}\,\alpha'_{ai}.
\end{equation}
Because the coefficients $\alpha'_{ai}$ are significantly sparser than those $\alpha_{ai}$ obtained through the standard Cholesky decomposition, the circuit required for preparing the approximate state $\rho'$ is considerably less expensive than that needed for preparing $\rho$. This offers an appealing approach to generating a high-quality approximate state, invoking highly economical preprocessing and resulting in highly efficient circuits.

The fact that $\rho'$ is close to $\rho$ can be quantified in terms of the trace distance between them as
\begin{eqnarray}\label{eq:D rho rho'}
D(\rho,\rho') &:=& \frac{1}{2}\|\rho-\rho'\|_* \nonumber\\
&\sim& \frac{1}{2}\|\rho-\rho'\|_F \nonumber\\
&\sim& \|\rho\|_F \times O(\epsilon)
\sim O(\epsilon),
\end{eqnarray}
where, for any matrix $A$ with $\sigma_i(A)$ being its singular values, $\|A\|_*:=\Tr\sqrt{A^\dag A}=\sum_i\sigma_i(A)$ is the trace norm (also known as the nuclear norm) of $A$, and $\|A\|_F:=\sqrt{\Tr (A^\dag A)}=\sqrt{\sum_i\sigma_i^2(A)}$ is the Frobenius norm (also known as the Hilbert--Schmidt norm).
The relation \eqref{ichol approx} with $M=\rho, M'\approx\rho'$ and the fact $\|\rho\|_F\leq \|\rho\|_*=1$ have been used.
Consequently, given a small drop tolerance $\epsilon$, the Fuchs--van de Graaf inequality \cite{761271}, $1-\sqrt{F(\rho_1,\rho_2)} \leq D(\rho_1,\rho_2) \leq \sqrt{1-F(\rho_1,\rho_2)}$, implies that the fidelity between $\rho$ and $\rho'$ remains as high as
\begin{equation}\label{fidelity}
1-O(\epsilon) \lesssim F(\rho,\rho') \lesssim 1-O(\epsilon^2).
\end{equation}

\section{Benchmarks}\label{sec:benchmark}

In this section, we demonstrate that the purified states obtained via the Cholesky decomposition are significantly sparser---and potentially require substantially fewer quantum gates for state-preparation circuits---than those obtained via the eigenvalue decomposition, particularly for highly sparse density matrices. Here the sparsity of an $n$-qubit pure state $\ket{\psi}=\sum_{a=0}^{2^n-1}\alpha_a\ket{a}$, where $\{\ket{a}\}$ denote the computational basis states, is defined as the number of nonzero coefficients $\alpha_a$ divided by $2^n$. When the resulting purified state satisfies the \emph{extremely sparse} condition, namely when the sparsity is smaller than $O(n/2^n)$, existing isometry-based methods tailored for extremely sparse isometries can be employed to construct such a purified state, thereby dramatically reducing the required gate count.

In addition, we show that the incomplete Cholesky decomposition can further increase the sparsity of the resulting purified states, even when only a very small drop tolerance is prescribed. As the drop tolerance increases, the sparsity of the purified state increases accordingly, leading to a reduction in the gate count of the corresponding circuit. This reduction can be substantial, while the gate fidelity is only slightly degraded when the drop tolerance remains relatively small. Unfortunately, with currently available algorithms, incomplete Cholesky solvers are not fully stable or scalable due to the dropping strategy. Consequently, our numerical tests are limited in scale and should be regarded primarily as a proof of concept.

For the classical preprocessing, both the Cholesky decomposition and the eigenvalue decomposition are carried out using the SciPy package.\footnote{In the SciPy package, the Cholesky decomposition does not directly support complex-valued matrices. Instead, we obtain the Cholesky factors indirectly from LU and LDL decompositions and select the sparser result.} We also utilize the random sparse matrix generator provided by SciPy to produce sampled density matrices with specified sparsity levels and ranks. Because the sparsity of the generated matrices cannot exactly match the target value due to algorithmic limitations,\footnote{In SciPy, a random $n\times r$ matrix $H$ can be generated with a specified sparsity level. A random square matrix of rank $r$ is then constructed as $H H^\dagger$. While increasing the sparsity of $H$ generally leads to greater sparsity in $H H^\dagger$, the precise relationship between the two is not straightforward.} we introduce a tolerance parameter $\delta_s$. If the difference between the desired and resulting sparsity lies within $\delta_s$, the generated density matrix is accepted; otherwise, a new sample is generated until the criterion is satisfied.

To test the complexity of the corresponding quantum circuits, we consider the isometry-based circuit construction for preparing the purified state (i.e., Methods A--E in Table~\ref{tab:gate_count}), instead of the method based on uniformly controlled rotations shown in \figref{fig:circuit for rho via cU}. Not only does the former generally yield more efficient circuits than the latter, but more importantly, the sparsity of the purified state can be further exploited by employing circuit synthesis methods tailored for sparse isometries (i.e., Methods D and E in Table~\ref{tab:gate_count}).

We first compute the theoretical upper bounds for Methods A, B, and C, as listed in Table~\ref{tab:gate_count}, to serve as baseline references. We then employ Method D to prepare the purified state obtained from the Cholesky decomposition, and Methods D and E to prepare the purified state obtained from the eigenvalue decomposition.\footnote{Note that Method~E does not apply to the Cholesky decomposition, as the method of \figref{fig:circuit for rho via isometry} requires $\rho$ to be expressed as an ensemble of \emph{orthonormal} pure states---a form that the Cholesky method generally does not produce.} In particular, we adopt the method proposed in \cite{Malvetti2021quantumcircuits}, which is specifically designed to exploit sparse isometries. The complexity of the synthesized circuits is then compared both with one another and with the theoretical bounds in order to quantify how effectively the sparsity of $\rho$ can be exploited.

\subsection{Cholesky decomposition versus eigenvalue decomposition}

\begin{figure}
\centering
\includegraphics[width=\textwidth]{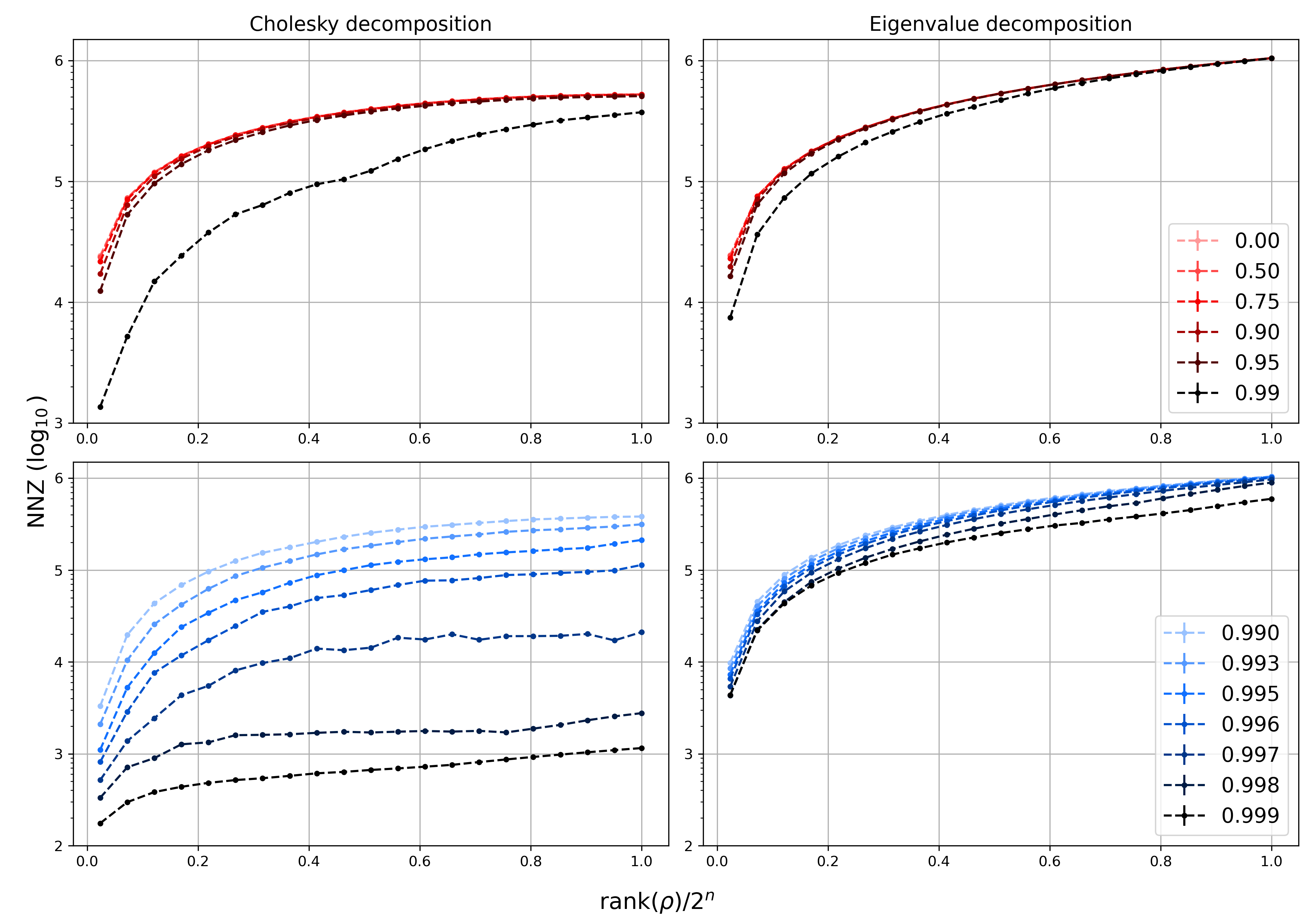}
%
\caption{Number of nonzero coefficients (NNZ) of the resulting purified states obtained from the Cholesky decomposition (left panels) and the eigenvalue decomposition (right panels) for 10-qubit density matrices. The upper panels show results for sparsity levels ranging from 0 to 0.99 with tolerance $\delta_s = 0.01$. The lower panels focus on extremely high sparsity levels ranging from 0.990 to 0.999 with tolerance $\delta_s = 0.0002$. In the upper panels, some data points for sparsity below 0.90 lie very close to one another and therefore appear nearly indistinguishable.}
\label{fig:sparsity10}
\end{figure}

We test random 10-qubit density matrices $\rho$ with different ranks and sparsity levels, where the sparsity is defined as the number of nonzero elements (NNZ) of $\rho$ divided by $2^n \times 2^n$. The number of nonzero coefficients (also denoted as NNZ) of the resulting purified states is shown in \figref{fig:sparsity10} for both the Cholesky decomposition (left panels) and the eigenvalue decomposition (right panels). Each data point represents the average over 50 randomly generated samples. The upper panels present results for low, moderate, and high sparsity levels ranging from 0.00 to 0.99, while the lower panels focus on extremely high sparsity levels ranging from 0.990 to 0.999. The rank of the density matrices varies from 21 to 1024 in 21 equally spaced increments.

The upper panels show that the Cholesky decomposition consistently yields purified states with fewer nonzero coefficients than those obtained from the eigenvalue decomposition, even when the density matrices $\rho$ exhibit low or moderate sparsity. Moreover, the advantage of the Cholesky decomposition becomes increasingly pronounced as the sparsity of $\rho$ increases. This trend is particularly evident in the lower panels, where the purified states obtained from the Cholesky decomposition are substantially sparser than those obtained from the eigenvalue decomposition.

\begin{figure}
\centering
\includegraphics[width=\linewidth]{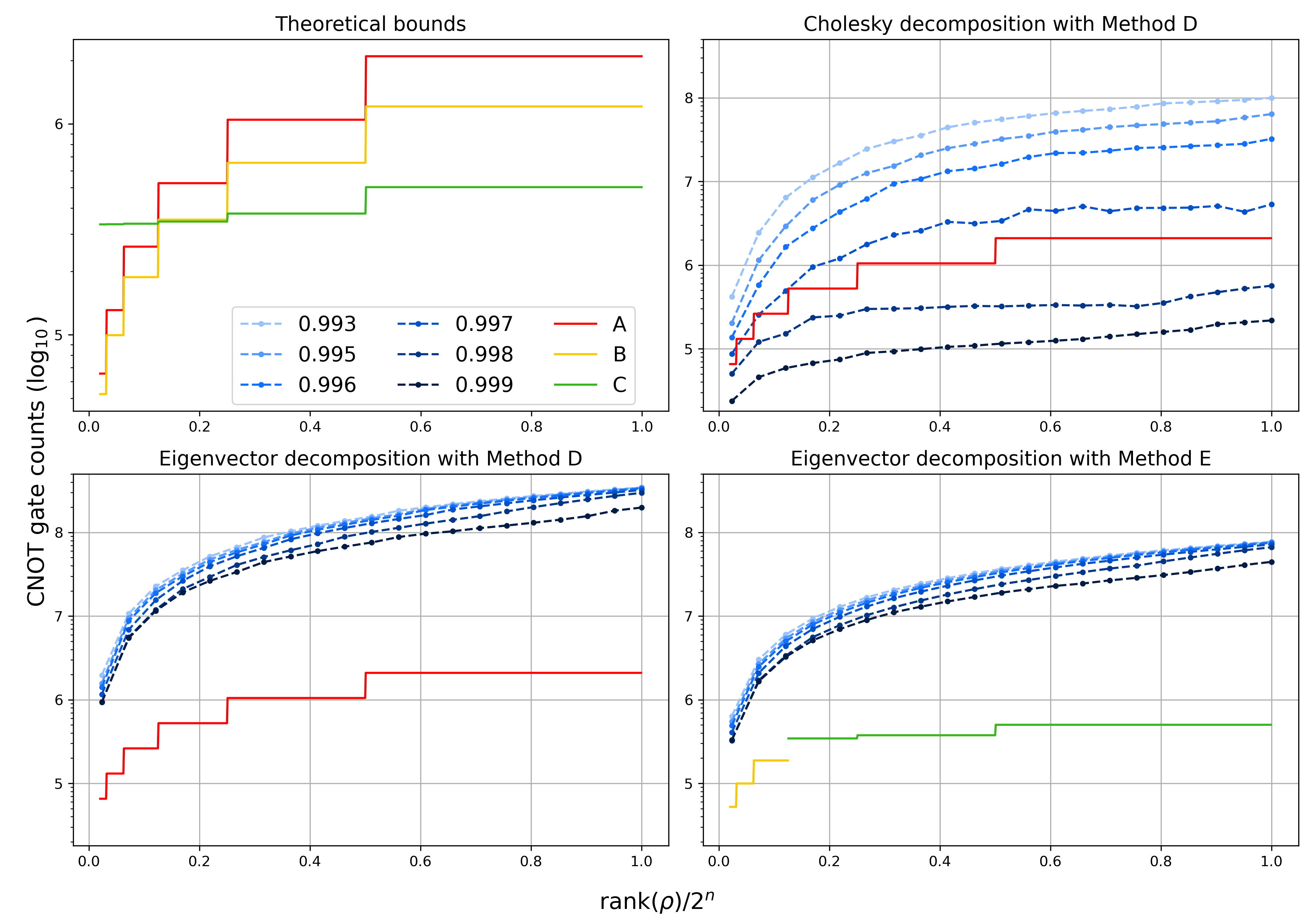}
\caption{CNOT gate counts of the synthesized circuits for preparing 10-qubit density matrices at extremely high sparsity levels. The upper-left panel shows the theoretical upper bounds of Methods~A, B, and~C in Table~\ref{tab:gate_count} as baseline references, while the remaining panels present the gate counts obtained using sparse-isometry synthesis methods.}
\label{fig:cx count q10}
\end{figure}

The complexities of the corresponding synthesized quantum circuits---measured in terms of the CNOT gate count---for the cases of extremely high sparsity are shown in \figref{fig:cx count q10}. The upper-left panel displays the theoretical upper bounds for Methods~A, B, and~C as baseline references. The upper-right panel presents the results obtained from the Cholesky decomposition using Method~D, compared with the upper bound of Method~A. The lower-left panel shows the results from the eigenvalue decomposition using Method~D, again referenced to the Method~A upper bound. The lower-right panel shows the results from the eigenvalue decomposition using Method~E, referenced to the smaller of the Method~A and Method~B upper bounds.

The Cholesky decomposition combined with Method~D performs significantly better than the eigenvalue decomposition combined with Method~D for all cases, and also outperforms the eigenvalue decomposition combined with Method~E whenever the sparsity of $\rho$ is sufficiently high. The eigenvalue-decomposition approaches with Methods~D and~E both yield CNOT gate counts larger than the upper bounds of the corresponding methods for constructing generic isometries (i.e., Methods~A, B, and~C). This indicates that the eigenvalue decomposition does not effectively exploit the (extreme) sparsity of the target matrix $\rho$, making it inefficient to employ circuit synthesis methods tailored for sparse isometries (i.e., Methods~D and~E). By contrast, the Cholesky decomposition not only exploits but often enhances sparsity. Consequently, employing Method~D, which is tailored for sparse isometries, can readily outperform Method~A, which is designed for generic isometries, once the target matrix $\rho$ becomes sufficiently sparse.

\begin{figure}
\centering
\includegraphics[width=\linewidth]{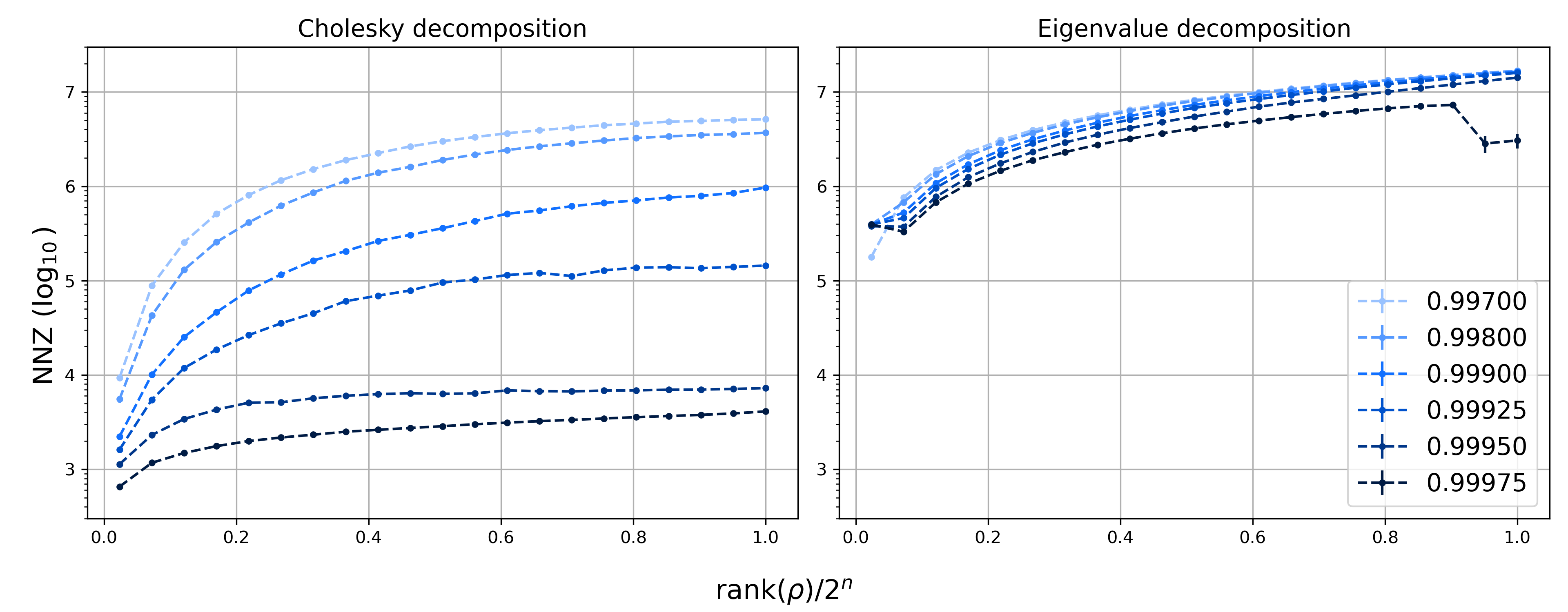}
\caption{Number of nonzero cofficients (NNZ) of the resulting purified states obtained from the Cholesky decomposition (left panel) and the eigenvalue decomposition (right panel) for 12-qubit density matrices at extremely high sparsity levels. The tolerance for the sampled density matrices is $\delta_s = 5\times 10^{-6}$. The kink points appearing in the eigenvalue-decomposition curves arise from numerical instability.}
\label{fig:sparsity12}
\end{figure}

\begin{figure}
\centering
\includegraphics[width=\linewidth]{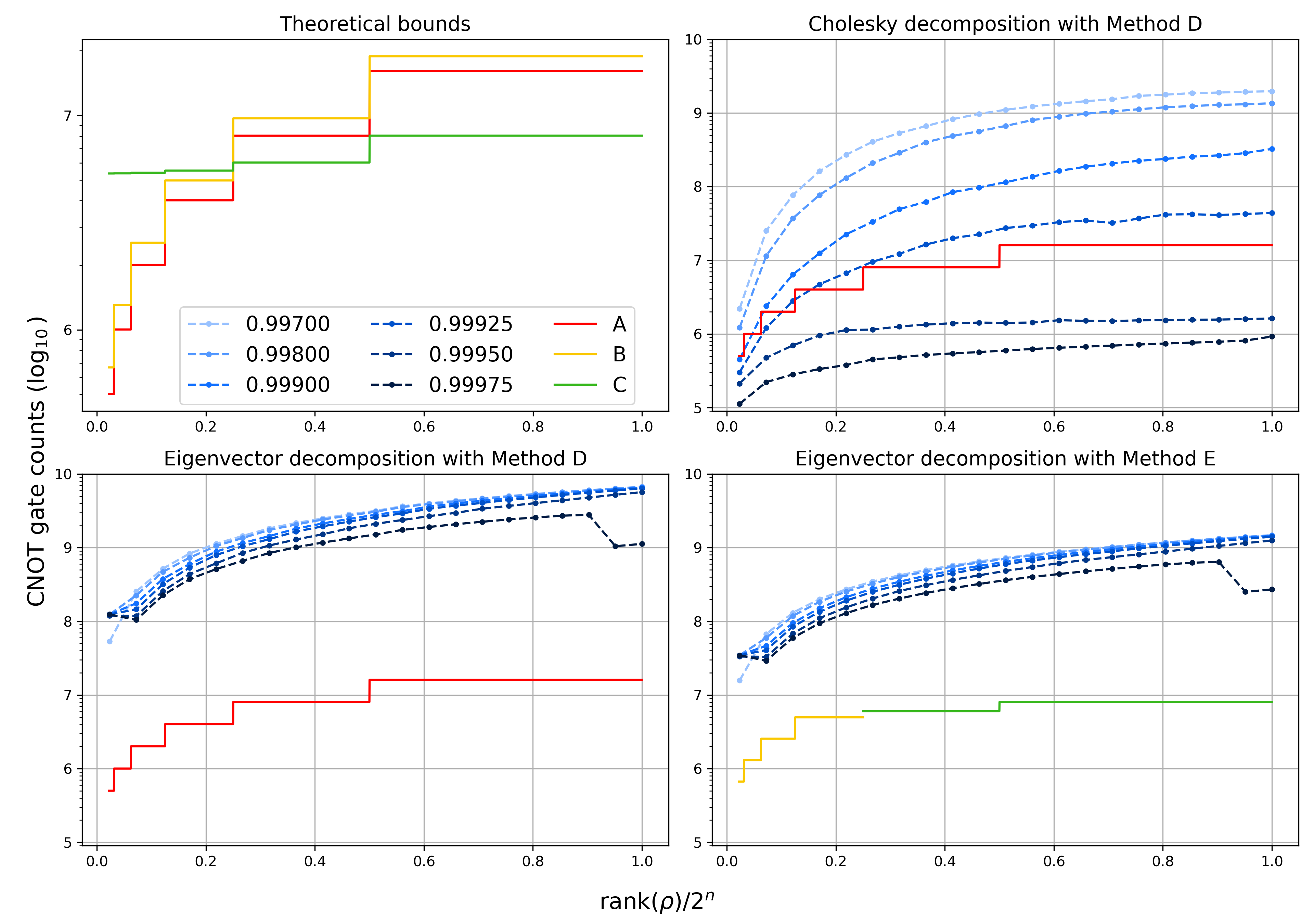}
\caption{CNOT gate counts of the synthesized circuits for preparing 12-qubit density matrices at extremely high sparsity levels. The kink points in the eigenvalue-decomposition curves are caused by numerical instability in the SciPy implementation.}
\label{fig:cx count q12}
\end{figure}

Similarly, we also test random 12-qubit density matrices at extremely high sparsity levels, ranging from 0.99700 to 0.99975, and present the results in \figref{fig:sparsity12} and \figref{fig:cx count q12}. These figures show the NNZ of the resulting purified states and the corresponding CNOT gate counts, respectively. The rank of the density matrices varies from 96 to 4096 in 21 equally spaced increments, and each data point represents the average over 20 samples. The results for the 12-qubit density matrices confirm the same advantage of the Cholesky decomposition observed in the 10-qubit cases.

Overall, our results demonstrate that the Cholesky decomposition consistently produces significantly sparser matrices---and consequently sparser purified states---than the eigenvalue decomposition, with the advantage becoming especially pronounced at extremely high sparsity levels. For density matrices in this regime, the increased sparsity obtained from the Cholesky decomposition leads to a substantial reduction in the CNOT gate count when existing algorithms for sparse isometries are applied. By contrast, the limited sparsity gains obtained from the eigenvalue decomposition provide little practical benefit for such methods.



\subsection{Incomplete Cholesky decomposition}\label{sec:benchmark incomplete Cholesky}

We further investigate how the incomplete Cholesky decomposition can increase the sparsity of the resulting purified state and how much the approximated state $\rho'$ deviates from the target state $\rho$ under varying drop tolerances. The sparsity of the resulting purified state is quantified by the number of nonzero coefficients (NNZ), while the deviation between $\rho$ and $\rho'$ is measured by the Frobenius norm $\|\rho-\rho'\|_F$. We consider full-rank\footnote{The incomplete Cholesky decomposition is implemented through the incomplete LU decomposition in the SciPy package, which currently supports only full-rank matrices.} yet sparse density matrices.

\begin{figure}
\centering
\includegraphics[width=.47\linewidth]{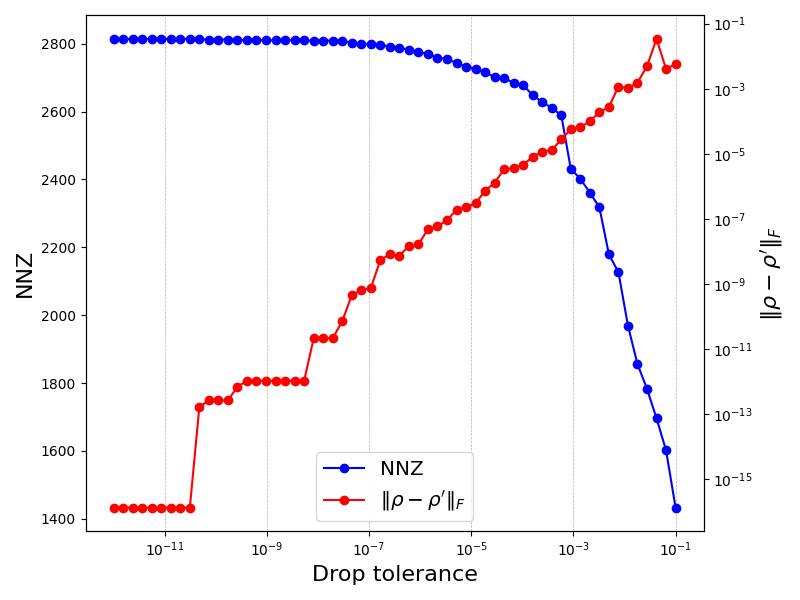}
\includegraphics[width=.47\linewidth]{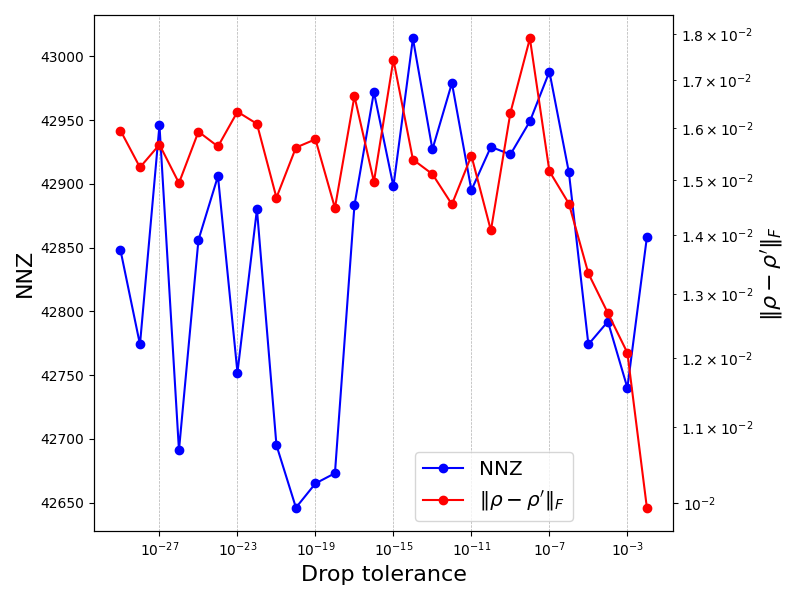}
\caption{Number of nonzero coefficients (NNZ) of the resulting purified state and the Frobenius norm $\|\rho-\rho'\|_F$ as functions of the drop tolerance in the incomplete Cholesky decomposition for a random density matrix of size $200\times200$ with sparsity $0.9673$ (left panel) and for a random density matrix of size $1024\times1024$ with sparsity $0.9908752$ (right panel).}
\label{fig:icho}
\end{figure}

First, we test a random density matrix of size $200\times200$ with sparsity $0.9673$, which yields $\text{NNZ}=6905$ when the complete Cholesky decomposition is used. The resulting NNZ and $\|\rho-\rho'\|_F$ obtained from the incomplete Cholesky decomposition with drop tolerance $\epsilon$ ranging from $10^{-12}$ to $10^{-1}$ are shown in the left panel of \figref{fig:icho}. The behavior of $\|\rho-\rho'\|_F$ as a function of $\epsilon$ agrees well with the theoretical estimate given in \eqref{eq:D rho rho'}. As expected, NNZ decreases as $\epsilon$ increases. Remarkably, even with a minuscule drop tolerance $\epsilon\lesssim3\times 10^{-11}$, NNZ can be substantially reduced from $\text{NNZ}=6905$ to approximately $\text{NNZ}\sim2800$, while the Frobenius norm remains extremely small, $\|\rho-\rho'\|_F\lesssim 1.4\times 10^{-16}$.\footnote{For comparison, numerical errors alone already produce $\|\rho-\rho'\|_F\sim10^{-17}$ even when the complete Cholesky decomposition is used.}

Next, we test a random density matrix of size $1024\times1024$ with sparsity $0.9908752$, which yields $\text{NNZ}=268332$ under the complete Cholesky decomposition. The corresponding results for NNZ and $\|\rho-\rho'\|_F$, obtained using the incomplete Cholesky decomposition with drop tolerance $\epsilon$ ranging from $10^{-29}$ to $10^{-2}$, are shown in the right panel of \figref{fig:icho}. In this case, the results deviate from the theoretical expectation and exhibit unstable behavior. This discrepancy arises because the currently available algorithm becomes numerically unstable when the drop-tolerance strategy is applied to density matrices. In particular, removing small-valued matrix elements can cause intermediate matrices to lose positive semidefiniteness, requiring additional modifications (fill-ins) to continue the Cholesky factorization. Moreover, for full-rank yet sparse density matrices, the off-diagonal elements are often several orders of magnitude smaller than the diagonal elements; dropping them can therefore significantly distort the factorization and degrade its accuracy. Despite these limitations, the incomplete Cholesky decomposition still reduces the NNZ dramatically from $\text{NNZ}=268332$ to approximately $\text{NNZ}\sim4300$.

These benchmarks demonstrate---as a proof of concept---the potential benefit of the incomplete Cholesky decomposition for efficiently approximating sparse density matrices. Unfortunately, existing algorithms for incomplete Cholesky decomposition are quite restrictive (e.g., limited to full-rank matrices), numerically unstable, and not scalable. Nevertheless, these limitations are not fundamental and could in principle be overcome. If a robust and scalable incomplete Cholesky solver becomes available, our proposal of approximating the target density matrix via incomplete Cholesky decomposition could provide substantial advantages.

\section{Summary}\label{sec:summary}

As a novel contribution to the literature, we explore the issue of arbitrary mixed state preparation. When the density matrix $\rho$ is specified as an ensemble of pure states, i.e., in the form of \eqref{eq:density matrix via state}, the circuit depicted in \figref{fig:circuit for rho via add} can be employed to generate $\rho$. This circuit necessitates the use of $2n+1$ qubit registers in the dynamic circuit framework and $\ell(n+1)-1$ qubit registers in the static circuit framework. In the worst-case scenario, the circuit requires $\ell(2^{n+1}-2n-2)$ CNOT gates, $\ell(2^{n+1}-1)-1$ one-qubit rotation gates, and $n(\ell-1)$ CSWAP gates in total.
Alternatively, the circuit presented in \figref{fig:circuit for rho via cU} can also generate $\rho$, bringing an additional benefit of providing a purified state of $\rho$, which could be useful in various situations. In the worst-case scenario, this circuit requires $2^m(2^{n+1}-1)-2n-1\sim\ell(2^{n+1}-1)-2n-1$ CNOT gates and $2^m(2^{n+1}-1)-1\sim\ell(2^{n+1}-1)-1$ one-qubit rotation gates, where $m=\lceil\log_2\ell\rceil$.
In the special case where the pure states $\ket{\psi_i}$ in the ensemble expression \eqref{eq:density matrix via state} are mutually orthogonal, the target density matrix $\rho$ can be prepared slightly more efficiently using the circuit shown in \figref{fig:circuit for rho via isometry}. The worst-case CNOT gate counts required by this circuit are summarized in Table~\ref{tab:gate_count}.

On the other hand, when the density matrix $\rho$ is specified by its matrix elements, as shown in \eqref{eq:density matrix component form}, one can perform the preprocessing of solving the eigenvalue problem to render \eqref{eq:density matrix component form} into the diagonal form of \eqref{eq:density matrix via eigenstates} before applying the aforementioned methods. However, solving the eigenvalue problem can be time-consuming, and the resulting eigenstates may be difficult to deal with.
Instead of solving the eigenvalue problem, we propose a better strategy that utilizes the Cholesky decomposition to factorize the $2^n\times 2^n$ density matrix $\rho$ into $\rho=AA^\dag$, where $A$ is a $2^n\times\ell$ matrix with $r\equiv\text{rank}(\rho)\leq\ell\leq2^n$.
Once $A$ is obtained, we can compute $p_i$ by \eqref{pi from A} and $\alpha_{ai}$ by \eqref{Aai}.
Consequently, by providing the states $\ket{\alpha_i}$ given by \eqref{eq: weight state} and \eqref{alpha i}, and $\rho_i=\ket{\psi_i}\bra{\psi_i}$ as the states $\ket{\psi_i}$ given by \eqref{psi expansion}, we can implement the circuit of \figref{fig:circuit for rho via add} to produce $\rho$.
Alternatively, we can implement the circuit of \figref{fig:circuit for rho via cU} or adopt other pure state preparation methods to produce the purified state $\ket{\Psi}^{(n+m)}$ given in \eqref{rho purification}, with the number of ancilla qubits selected as $m=\lceil\log_2\ell\rceil$.

For a $d\times d$ matrix $M$, the commonly used algorithms for solving the Cholesky decomposition have a time complexity of $O(d^3)$ and require $\sim d^3/3$ floating-point operations, in general more efficient than the commonly used QR algorithm for solving the eigenvalue problem, which also has a time complexity of $O(d^3)$ but requires $\sim 6d^3$ floating-point operations.
In cases where the rank of $M$ is sufficiently low, the application of rank-aware algorithms for the Cholesky decomposition can significantly enhance computational efficiency. Additionally, if $M$ is sparse enough, computational efficiency can be further improved and a sparse Cholesky factorization can be obtained by employing reordering strategies.

The resulting matrix $A$ typically exhibits a considerable degree of sparsity, even when the density matrix $\rho$ itself is not sparse at all. Particularly, in the case where $\rho$ is sparse enough, reordering strategies can be employed to minimize fill-ins and thus yield high sparsity in the resulting $A$. This sparsity in the matrix $A$ offers a remarkable advantage in circuit implementation.
Given that many of the coefficients $\alpha_{ai}$ vanish, the corresponding states $\ket{\psi_i}$ and the purified state $\ket{\Psi}^{(n+m)}$ in \eqref{rho purification} too are sparse in their computational basis. Consequently, by adopting the methods specially designed for the preparation of sparse pure states, such as those proposed in \cite{Malvetti2021quantumcircuits, 9586240, de2022double, PhysRevA.106.022617}, the complexity of the circuit for preparing $\rho$ can be significantly reduced.

Furthermore, the incomplete Cholesky decomposition with the threshold dropping option provides an appealing scheme for preparing a high-fidelity approximate state $\rho'$ close to $\rho$. Striking an excellent balance between accuracy and efficiency, this approach incurs an extremely low computational cost in preprocessing and yields highly efficient circuits in the end. When the drop tolerance $\epsilon$ increases, the sparsity of the Cholesky factorization rises, while the accuracy of the approximation declines. Consequently, with increasing drop tolerance, both the efficiency of the preprocessing and the efficiency of the resulting quantum circuit improve significantly, while the fidelity between $\rho$ and $\rho'$ degrades but remains satisfactorily high, as indicated in \eqref{fidelity}.

In addition to the theoretical analyses, we perform performance benchmarks for methods based on complete and incomplete Cholesky decompositions. Our results show that the Cholesky decomposition consistently produces purified states with far fewer nonzero coefficients than those obtained from the eigenvalue decomposition. Consequently, when existing isometry-based synthesis techniques designed for extremely sparse isometries are applied, the resulting circuits exhibit substantial reductions in complexity. Moreover, because the Cholesky decomposition significantly enhances sparsity, it may also offer advantages for preparing density matrices with low or moderate sparsity. Such cases remain relevant for small-scale systems (or small subsystems within larger ones), and these potential benefits could be realized if circuit-synthesis methods tailored for highly---but not necessarily extremely---sparse isometries are developed.\footnote{Existing circuit-synthesis methods that exploit sparsity in quantum states primarily target extremely sparse cases. In principle, analogous methods tailored for highly---but not necessarily extremely---sparse states could also be developed.}

Our benchmarks on the incomplete Cholesky decomposition further provide a proof of concept that sparsity can be significantly increased even when only a very small drop tolerance is used. This approach enables high-fidelity approximations of the target density matrix while incurring only mild fidelity loss, thereby offering an additional pathway to reducing circuit complexity. Although currently available incomplete Cholesky solvers remain restrictive, numerically unstable, and not yet scalable, these limitations arise from algorithmic implementations rather than fundamental barriers and could in principle be overcome with improved numerical techniques.

\begin{acknowledgments}
This work was supported in part by the National Science and Technology Council, Taiwan, under Grant Nos.\ 112-2119-M-002-017, 113-2119-M-002-024, 114-2119-M-002-020, 114-2811-M-002-034, 115-2119-M-002-006 and 115-2811-M-002-039; by the NTU Center of Data Intelligence: Technologies, Applications, and Systems under Grant Nos.\ NTU-113L900903 and NTU-114L900903; by the NTU Core Consortium Project NTU-CC114L895002; and by the Intelligent Computing for Sustainable Development Research Center at National Taiwan Normal University under the Higher Education Sprout Project of the Ministry of Education, Taiwan.
\end{acknowledgments}

\appendix

\section{Preparation of arbitrary pure states}\label{sec:arbitrary state}
Here, we present a circuit synthesis scheme for preparing arbitrary pure states from our own perspective. By optimizing this circuit with uniformly controlled one-qubit gates, as described in \appref{sec:uniformly control}, we obtain a construction that is ultimately equivalent to the one proposed in \cite{PhysRevA.71.052330, mottonen2004transformatio}.

Any arbitrary $n$-qubit quantum state can be described as $\ket{\psi}^{(n)}=\sum_{a=0}^{2^n-1}\alpha_a\ket{a}$, where $\ket{a}$ are the computational basis states.
Given the coefficients $\alpha_a\in\mathbb{C}$ all specified, the state $\ket{\psi}^{(n)}$ can be produced by the following systematic scheme.

In the case of $n=1$, any arbitrary state
\begin{eqnarray}\label{psi n=1}
\ket{\psi}^{(n=1)}&=&\alpha\ket{0}+\beta\ket{1} \nonumber\\
&\equiv& e^{-i\phi/2}\cos\frac{\theta}{2}\ket{0}-e^{i\phi/2}\sin\frac{\theta}{2}\ket{1}
\end{eqnarray}
can be produced by the circuit in \figref{fig:circuit for 1-qubit state}.

\begin{figure}
\begin{quantikz}
\lstick{$\ket{0}$} & \gate{R_y(\theta)} & \gate{R_z(\phi)} & \rstick{$\ket{\psi}^{(n=1)}$} \qw
\end{quantikz}
\caption{The quantum circuit producing the state in \eqref{psi n=1}.}
\label{fig:circuit for 1-qubit state}
\end{figure}

For $n=k$, suppose that we can always construct a circuit that maps the state $\ket{0}^{\otimes k}$ to a desired arbitrary $k$-qubit state $\ket{\psi}^{(k)}$. Denote this circuit as a unitary gate $U[\ket{\psi}^{(k)}]$.

For $n=k+1$, then, any arbitrary $(k+1)$-qubit state can be expressed as
\begin{eqnarray}\label{psi n=k+1}
\ket{\psi}^{(n=k+1)}=\alpha\ket{0}\ket{\psi_1}^{(k)}+\beta\ket{1}\ket{\psi_2}^{(k)}.
\end{eqnarray}
By induction, the state \eqref{psi n=k+1} can be produced by the circuit in \figref{fig:circuit for (k+1)-qubit state}.
If \eqref{psi n=k+1} takes the special form
\begin{eqnarray}\label{psi special}
\ket{\psi}^{(n=k+1)}=(\alpha\ket{0}+\beta\ket{1})\ket{\psi}^{(k)},
\end{eqnarray}
the first qubit is not entangled with the remaining $k$ qubits. In this case, the circuit in \figref{fig:circuit for (k+1)-qubit state} can be much simplified by disconnecting the controls from the first qubit and replacing $U[\ket{\psi_1}^{(k)}]$ and $U[\ket{\psi_2}^{(k)}]$ with a single gate $U[\ket{\psi}^{(k)}]$.

\begin{figure}
\begin{quantikz}
\lstick{$\ket{0}$} & \gate{R_y(\theta)} & \gate{R_z(\phi)} & \octrl{1} & \ctrl{1} & \qw\rstick[wires=2]{$\ket{\psi}^{(k+1)}$} \\
\lstick{$\ket{0}^{(k)}$} & \qwbundle{k} & \qw & \gate{U[\ket{\psi_1}^{(k)}]} \qw & \gate{U[\ket{\psi_2}^{(k)}]} \qw & \qw
\end{quantikz}
\caption{The quantum circuit producing the state in \eqref{psi n=k+1}.}
\label{fig:circuit for (k+1)-qubit state}
\end{figure}

By expanding the nested structure in \figref{fig:circuit for (k+1)-qubit state}, it is easy to see that the circuit consists of numerous controlled-$R_y$ and controlled-$R_z$ gates, conditioned on all different control node configurations. By exploiting the fact that controlled-gates conditioned on distinct control node configurations commute with one another, the sequence of the controlled-$R_y$ and controlled-$R_z$ gates can be rearranged into the equivalent circuit as illustrated in \figref{fig:circuit for n-qubit state} in terms of $k$-fold \emph{uniformly controlled rotations}, $F^k[R_\mathbf{a}]$, for $k=0,\dots,n-1$. Uniformly controlled rotations belong to a special case of \emph{uniformly controlled one-qubit gates}, which were first introduced in \cite{PhysRevLett.93.130502} and investigated in \cite{PhysRevLett.93.130502,PhysRevA.71.052330,mottonen2004transformatio}.\footnote{For the convenience of readers, a brief outline of uniformly controlled one-qubit gates is provided in \appref{sec:uniformly control}.}
A $k$-fold uniformly controlled one-qubit gate $F^k[U]$ is a $(k+1)$-bit gate with one target qubit and $k$ control qubits in the form of \eqref{uniformly controlled gate}.
The circuit in \figref{fig:circuit for n-qubit state} is exactly the same as the state preparation circuit proposed in \cite{PhysRevA.71.052330, mottonen2004transformatio}, albeit derived from a slightly different perspective.

According to the strategy proposed in \cite{PhysRevA.71.052330,mottonen2004transformatio}, a $k$-fold uniformly controlled rotation $F^k[R_\mathbf{a}]$ can be efficiently decomposed into $2^k$ CNOT gates and $2^k$ one-qubit rotation gates, as illustrated in \figref{fig:uniformly controlled rotation}. Furthermore, before concatenating $F^k[R_y]$ and $F^k[R_z]$, we can equivalently implement $F^k[R_z]$ by applying each gate as shown in \figref{fig:uniformly controlled rotation} but in the opposite sequential order, while replacing each rotation angle $\phi_a$ with $-\phi_a$. Then, for each pair of $F^k[R_y]$ and $F^k[R_z]$, the rightmost CNOT gate of $F^k[R_y]$ and the leftmost CNOT gate of $F^k[R_z]$ can be cancelled out, further reducing two CNOT gates. As a result, the quantum circuit in \figref{fig:circuit for n-qubit state} requires $\sum_{k=0}^{n-1}(2\times2^{k}-2)=2^{n+1}-2n-2$ CNOT gates and $\sum_{k=0}^{n-1}2\times2^{k}=2^{n+1}-2$ one-qubit rotation gates in total.
The requirement of $2^{n+1}-2$ one-qubit rotation gates corresponds to $2^n-1$ real degrees of freedom for arbitrary $n$-qubit states.

\begin{figure}
\begin{quantikz}[column sep=0.25cm]
\lstick{$\ket{0}^{(n)}$}
& \gate{F^0[R_y]}\qwbundle{n}
& \gate{F^0[R_z]}
& \gate{F^1[R_y]}
& \gate{F^1[R_z]}
& \ \ldots \
& \gate{F^{n-1}[R_y]}
& \gate{F^{n-1}[R_z]}
&\rstick{$\ket{\psi}^{(n)}$}\\
\lstick[wires=5]{$\ket{0}^{(n)}$}
&\gate{R_y}
&\gate{R_z}
&\uctrl{1}
&\uctrl{1}
&\ \ldots\
&\uctrl{1}
&\uctrl{1}
& \rstick[wires=5]{$\ket{\psi}^{(n)}$}\\
&&
&\gate{R_y}
&\gate{R_z}
&\qw\ \ldots\
&\uctrl{1}
&\uctrl{1}&\\
\setwiretype{n}&&&&&& \vdots&\vdots&\\
&&&&&\ \ldots\ & \uctrl{1}\vqw{-1}&\uctrl{1}\vqw{-1}&\\
&&&&&\ \ldots\ &\gate{R_y}&\gate{R_z}&\\
\end{quantikz}
\caption{The quantum circuit equivalent to \figref{fig:circuit for (k+1)-qubit state}. The upper plot is a concise representation of the lower one. A rectangle of $R_\mathbf{a}$ along with the vertical line adorned with $k$ split dots represents a $k$-fold \emph{uniformly controlled rotation} $F^k[R_\mathbf{a}]$ (see \appref{sec:uniformly control} for more detail).}
\label{fig:circuit for n-qubit state}
\end{figure}

\section{Uniformly controlled one-qubit gates}\label{sec:uniformly control}
In this appendix, we provide a brief outline of \emph{uniformly controlled one-qubit gates} and \emph{uniformly controlled rotations} as a special category. Readers are referred to \cite{PhysRevLett.93.130502,PhysRevA.71.052330, mottonen2004transformatio} for more details and explanations.

A $k$-fold uniformly controlled one-qubit gate $F^k[U]$ is a $(k+1)$-qubit gate with one target qubit and $k$ control qubits in the form
\begin{equation}\label{uniformly controlled gate}
F^k[U] := \bigg\{
\begin{array}{ll}
\sum_{a=0}^{2^k-1} \ket{a}\bra{a}\otimes U_a &\,\text{for}\quad k>0,\\
U_0&\, \text{for}\quad k=0,
\end{array}
\end{equation}
where $U_a$ are one-qubit unitary operators acting on the target qubit.
A uniformly controlled one-qubit gate $F^k[U]$ can be understood as a sequence of $k$-fold controlled gates conditioned on all different control node configurations as illustrated in \figref{fig:uniformly controlled}.

\begin{figure}
\begin{quantikz}[column sep=0.1cm]
&\uctrl{1}&\\
&\uctrl{1}&\\
&\gate{U}&
\end{quantikz}
=
\begin{quantikz}
\qw&\octrl{1}&\octrl{1}&\ctrl{1}&\ctrl{1}&\qw\\
\qw&\octrl{1}&\ctrl{1}&\octrl{1}&\ctrl{1}&\qw\\
\qw&\gate{U_0}&\gate{U_1}&\gate{U_2}&\gate{U_3}&\qw
\end{quantikz}
\caption{The quantum circuit of a uniformly controlled one-qubit gate $F^2[U]$.}
\label{fig:uniformly controlled}
\end{figure}

In the special case where all the one-qubit operators $U_a$ belong to the one-parameter group of rotations
about a common axis $\mathbf{a}$ that is perpendicular to the $x$-axis, i.e.\ $U_a=R_\mathbf{a} (\theta_a)$ with $\mathbf{a}\cdot\hat{x}=0$ and some angle $\theta_a$, the uniformly controlled gate $F^k[U]$ is called a \emph{uniformly controlled rotation} and denoted as $F^k[R_\mathbf{a}]$.

Efficient implementation of uniformly controlled one-qubit gates can be achieved in terms of CNOT gates and one-qubit gates. Specifically, a $k$-fold uniformly controlled rotation can be implemented even more efficiently with $2^k$ CNOT gates and $2^k$ one-qubit rotation gates \cite{PhysRevLett.93.130502,PhysRevA.71.052330, mottonen2004transformatio} as illustrated in \figref{fig:uniformly controlled rotation}.

Given
\begin{equation}\label{uniformly controlled rotation}
F^k[R_\mathbf{a}] := \bigg\{
\begin{array}{ll}
\sum_{a=0}^{2^k-1} \ket{a}\bra{a}\otimes R_\mathbf{a}(\theta_a) &\,\text{for}\quad k>0,\\
R_\mathbf{a}(\theta_0)&\, \text{for}\quad k=0,
\end{array}
\end{equation}
the corresponding rotation angles $\phi_a$ of $R_\mathbf{a} (\phi_a)$ in \figref{fig:uniformly controlled rotation} are given via the transformation
\begin{equation}\label{gray code angle}
\phi_a = \sum_{b=0}^{2^k-1} M_{ab}\, \theta_b , \quad M_{ab} =2^{-k} (-1)^{\mathpzc{b}(b)\cdot\mathpzc{g}(a)},
\end{equation}
where $\mathpzc{b}(b)$ represents the binary code of integer $b$, $\mathpzc{g}(a)$ represents the binary reflected Gray code of integer $a$, and $\mathpzc{b}(b)\cdot\mathpzc{g}(a)$ is the bit-wise multiplication of $\mathpzc{b}(b)$ and $\mathpzc{g}(a)$ \cite{PhysRevLett.93.130502,PhysRevA.71.052330}.

\begin{figure}
\scalebox{0.85}{
\begin{quantikz}[column sep=0.1cm]
&\uctrl{1}&\\
&\uctrl{1}&\\
&\uctrl{1}&\\
&\gate{R_\mathbf{a}}&
\end{quantikz}
}
$\cong$
\scalebox{0.85}{
\begin{quantikz}[column sep=0.1cm]
\qw&\qw&\qw     &\qw&\qw     &\qw&\qw     &\qw&\ctrl{3}&\qw&\qw     &\qw&\qw     &\qw&\qw     &\qw&\ctrl{3}&\qw\\
\qw&\qw&\qw     &\qw&\ctrl{2}&\qw&\qw     &\qw&\qw     &\qw&\qw     &\qw&\ctrl{2}&\qw&\qw     &\qw&\qw     &\qw\\
\qw&\qw&\ctrl{1}&\qw&\qw     &\qw&\ctrl{1}&\qw&\qw     &\qw&\ctrl{1}&\qw&\qw     &\qw&\ctrl{1}&\qw&\qw     &\qw\\
\qw&\gate{R_\mathbf{a} (\phi_0)}&\targ{}&\gate{R_\mathbf{a} (\phi_1)}&\targ{}&\gate{R_\mathbf{a} (\phi_2)}&\targ{}&\gate{R_\mathbf{a} (\phi_3)}&\targ{}&\gate{R_\mathbf{a} (\phi_4)}&\targ{}&\gate{R_\mathbf{a} (\phi_5)}&\targ{}&\gate{R_\mathbf{a} (\phi_6)}&\targ{}&\gate{R_\mathbf{a} (\phi_7)}&\targ{}&\qw
\end{quantikz}
}
\caption{An efficient quantum circuit implementation for a uniformly controlled rotation $F^3[R_\mathbf{a}]$.}
\label{fig:uniformly controlled rotation}
\end{figure}

\bibliography{ref}

\end{document}